\documentclass[aps,pra,superscriptaddress,twocolumn,floatfix]{revtex4-1}
\usepackage[utf8]{inputenc}
\usepackage[T1]{fontenc}
\usepackage{amsmath}
\usepackage{amssymb}
\usepackage{mathtools}
\usepackage{graphicx}
\usepackage{overpic}
\usepackage[usenames,dvipsnames]{color}
\usepackage{bm}
\usepackage[colorlinks,urlcolor=blue,citecolor=blue,linkcolor=blue]{hyperref}
\usepackage{url}
\usepackage{subfigure}
\usepackage{slashed,bbm}
\usepackage{dsfont}
\usepackage{setspace}
\usepackage{soul}
\usepackage{braket}
\usepackage{sidecap}
\usepackage{hyperref}
\usepackage{multirow}

\begin{document}

%\allowdisplaybreaks
%-----------------------------------
\title{Spin susceptibilities in magnetic type-I and type-II Weyl semimetals}
%-----------------------------------

\author{Feng Xiong}
\author{Xingjie Han}
\author{Carsten Honerkamp}
\affiliation{Institut für Theoretische Festkörperphysik, RWTH Aachen University,
and JARA Fundamentals of Future Information Technology, Germany}

%-----------------------------------------
\date{\today}

\begin{abstract}
We investigate interacting spin susceptibilities in lattice models for $\mathcal{T}$-reversal symmetry-broken Weyl semimetals.
We employ a random phase approximation (RPA) method for the spin-SU(2)-symmetry-broken case that includes mixtures of ladder and bubble diagrams, beyond a SU(2)-symmetric case. Within this approach, the relations between the tendency towards magnetic order and the band structure tilt parameter $\gamma$ under different temperatures are explored. 
The critical interaction strength $U_c$ for magnetic ordering decreases as the tilt term changes from type-I Weyl semimetals to type-II. The lower temperature, the sharper is the drop in $U_c$ at the critical point between them. 
The variation of $U_c$ with a slight doping near half-filling is also studied. 
It is generally found that these Weyl systems show a strongly anisotropic spin response with an enhanced  doubly degenerate transverse susceptibility perpendicular to tilt direction, inherited from $\mathcal{C}_{4z}$ rational symmetry of bare Hamiltonian, but with the longitudinal response suppressed with respect to that. For small tilts $\gamma$ and strong enough interaction, we find two degenerate ordering patterns with spin order orthogonal to the tilt direction but much shorter spin correlation length parallel to the spin direction.
With increasing the tilt, the system develops instabilities with respect to in-plane magnetic orders with wavevector $(0,\pi, q_z)$ and $(\pi,0, q_z)$, with $q_z$ increasing from 0 to $\pi$ before the transition to a type-II Weyl semimetal is reached. 
These results indicate a greater richness of magnetic phases in correlated Weyl semimetals that also pose challenges for precise theoretical descriptions.
\end{abstract}
%-----------------------------------------

%-----------------------------------------
\maketitle
%-----------------------------------------
\section{INTRODUCTION}

Weyl semimetals (WSMs) \cite{murakami2007, XGWAN2011, Vishwanath2015, YanFelser2017, Armitage2018} have been intensively studied both experimentally and theoretically in recent years.
Similar to Dirac semimetals (DSMs) \cite{Liuzk2014, Yang2014, Young2015, Armitage2018}, they host linear dispersion near the bulk band crossing points, but break either inversion or time-reversal symmetry (TRS).
Moreover, the projections of bulk gapless Weyl nodes with opposite chiralities are connected by the well-known Fermi arc \cite{XGWAN2011}, which has been successfully observed experimentally in TaAs \cite{Lv2015,yang2015Weyl} and TaP \cite{Xu2015} via angle-resolved photoemission spectroscopy (ARPES). 
Subsequently, Soluyanov et al.\,\cite{Soluyanov2015} have extended the concept of WSM to type-I and type-II based on the inequivalent topology of Fermi surface.
Under controllable ways of doping or strain, the symmetric conical spectrum can be tilted along a certain direction in three-dimensional (3D) momentum space and a Lifshitz transition will happen between type-I and type-II. 
At small tilt, the Weyl node Fermi surface (FS) remains  and that is so-called type-I WSM.
If the tilt is strong, the conical spectrum will be tipped over and become type-II WSM with a pair of electron- and hole-pocket FS near a Weyl node.
Since the two kinds of Fermi surface cannot be adiabatically deformed, it leads to remarkable differences in physical properties including the chiral anomaly in Landau levels \cite{ChaoXing2013, Huang2015, Udagawa2016, Jia2016, Zhang2016} and anomalous hall or spin hall conductivity \cite{KaiYu2011, Burkov2014, ChaoXing2016, Liu2018, Peng2019, Menon2020, Singh2020, Garcia2020}.
Type-II WSM has also been studied in realistic materials including WTe$_2$ \cite{Soluyanov2015}, MoTe$_{2}$ \cite{Yan2015,Deng2016}, Mo$_{x}$W$_{1-x}$Te$_{2}$ \cite{Belopolski2016},  XP$_{2}$ (X=Mo, W) \cite{Yao2019}, Ta$_{3}$S$_2$ \cite{Change2016}, YbMnBi$_2$ \cite{Borisenko2016} and LaAlGe \cite{XuSu2017}.

The previous studies on WSMs can be well explained by the theory of non-interaction energy band.
More recently, experimental discoveries of superconductivity \cite{Qi2016, Kimeaao2018, Xing2019}, magnetism \cite{Kuroda2017, YangRun2020, Liu2020, Destraz2020}, and charge density wave \cite{Gooth2019, Shi2021} in WSMs necessitate accounting for electron-electron correlation effects .
However, it remains comparatively less explored in the field of correlated topological WSMs concerning experiments and theories.
From the theoretical side, renormalization group (RG) method \cite{Maciejko2014, Hsin2015, ShaoKai2015, Shixin2017, Lee2017, JingRong2019} has been employed to study low-energy effective Hamiltonian for an isolated Weyl node, as well as the Hartree-Fock (HF) mean-field approximation \cite{Xue2017}.
In contrast, fewer numerical methods were employed in correlated WSM lattice Hamiltonian, although this could reveal additional, important effects.
Among them, one aspect is that the HF method and cluster perturbation theory (CPT) have studied how interactions move and renormalize Weyl fermions \cite{William2012, Yixiang2017}. 
The others are that dynamical mean-field theory (DMFT) has been applied in multiple WSM systems \cite{Ara2012, Ivanov2019, Irsigler2020, Kundu2021}, i.e., the 3D pyrochlore iridates, muti-Weyl semimetals, and type-I or type-II WSM without inversion symmetry.
Yet, so far there is no coherent picture and approach to interaction effects in topological semimetals.

Until now, functional renormalization group methods for lattice models have become a widely used tool for the exploration of interaction effects in two-dimensional systems \cite{Metzner2012,Platt2012,QHWang2012,Lichtenstein2017}, with extension to three-dimensional cases \cite{Ehrlich2020}.
It is natural to extend it to topological band structures as well.
In this regard, it is helpful to first study ordinary many-body perturbation approaches in such models.
This gives insights on the main physical effects and also on technical or numerical issues.
In this paper, we study a TRS broken WSM with Hubbard interaction.
The authors of Ref. \cite{Soluyanov2015} already pointed out that the nonzero density of states of type-II could lead to new behavior when the Fermi energy lies at the Weyl node. 
For interaction-driven phenomena like ordered states, the density of states near the Fermi level as well as the shape of the Fermi surfaces are decisive parameters.
Therefore, one should expect that the correlated physics in terms of ordering and collective excitation will be different in the two cases and that the parameter-driven Lifshitz transition from type-I and type-II will witness clear changes in these properties.
The main goal of this paper is to work out whether this idea holds for a simple toy model containing two Weyl nodes that can be tuned from type-I to type-II.
By using the full scope of the RPA method, we find the amplitude of critical Hubbard interaction decreases with more tilt. Specifically, when slightly doping around Weyl-node Fermi energy, the correspondent critical interaction values behave differently between type-I and type-II, which indicates the sensitivity connected to the topographies of Fermi surfaces. Additionally, the exotic in-plane spin density orders are favorable in the correlated spin-orbit coupling (SOC) WSMs. With an additional in-plane rotational symmetry, the doubly degenerate planar spin texture will have integer or fraction spatial modulation along the tilt direction.
The emergent multi-$\bm{k}$  and incommensurate magnetism has been observed in a proposed WSM material CeAlGe \cite{Puphal2020} and the in-plane symmetry measurement setup has been realized in the experiment \cite{Hodovanets2021}.

This paper is organized as follows. Section \ref{Model} introduces the WSM Hamiltonian model and Hubbard interaction. In Sec. \ref{Method}, we derive the susceptibilities formula in the RPA method extended to a non-SU(2) scheme. The numerical results and analysis of susceptibilities in WSM are contained in Sec. \ref{Result}. Finally, Sec. \ref{Summary} contains concluding remarks.

\begin{figure*}[t]
\begin{overpic}[width=0.24\linewidth]{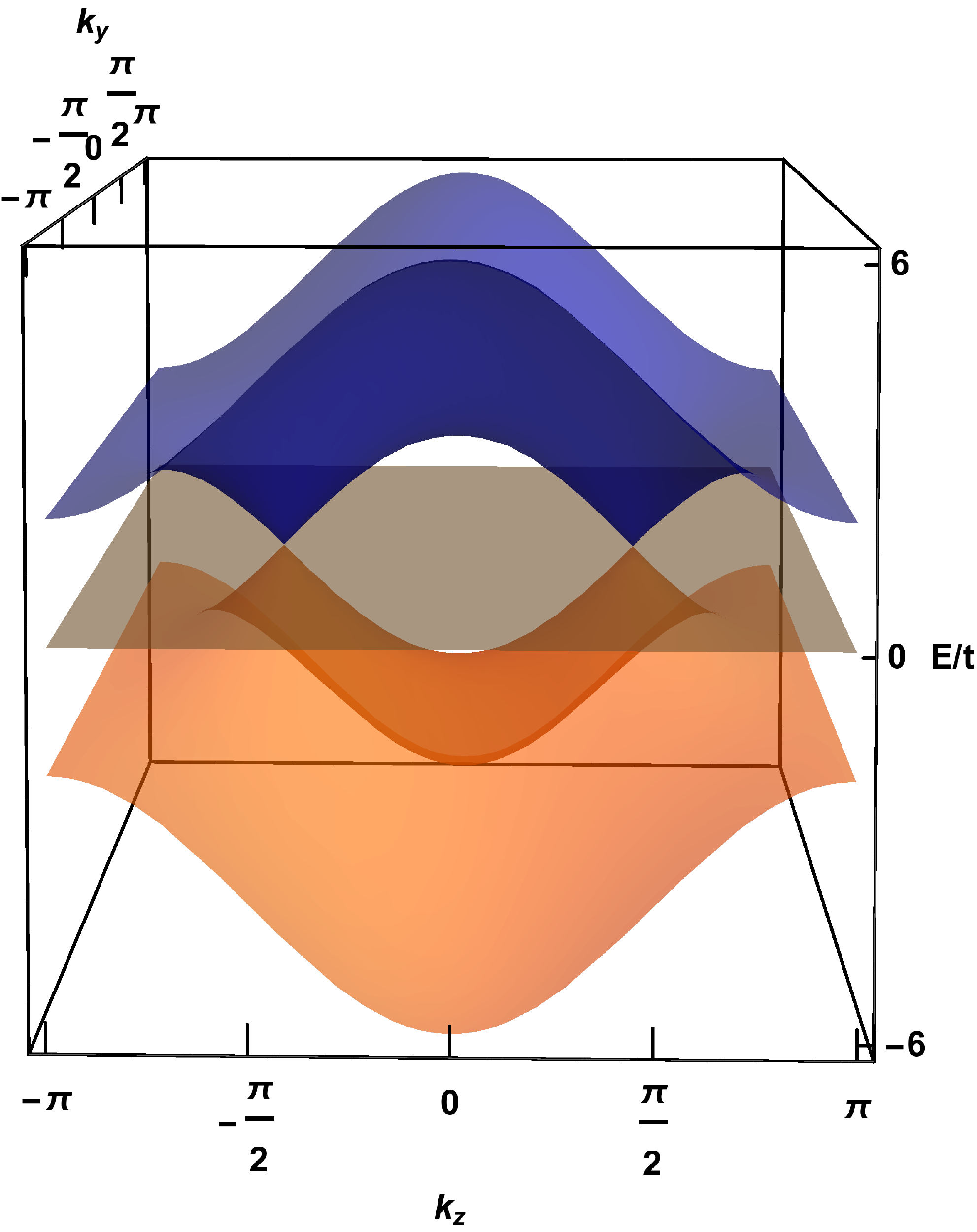}
\put(31,90){\begin{large} \bfseries $(\bf {a} )$ \end{large}}
\end{overpic}
\begin{overpic}[width=0.24\linewidth]{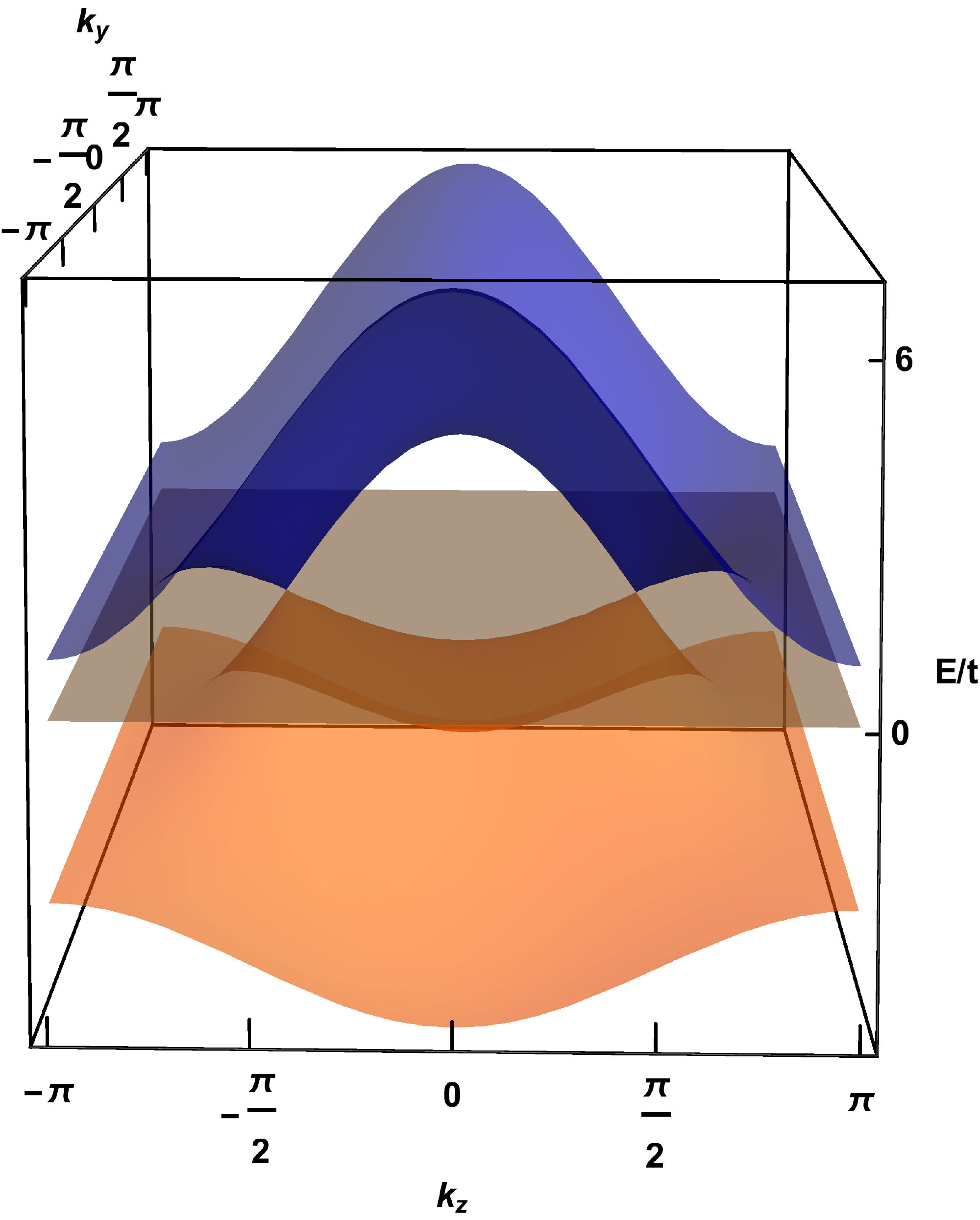}
\put(31,91){\begin{large} \bfseries $( \bf{b} )$ \end{large}}
\end{overpic}
\begin{overpic}[width=0.24\linewidth]{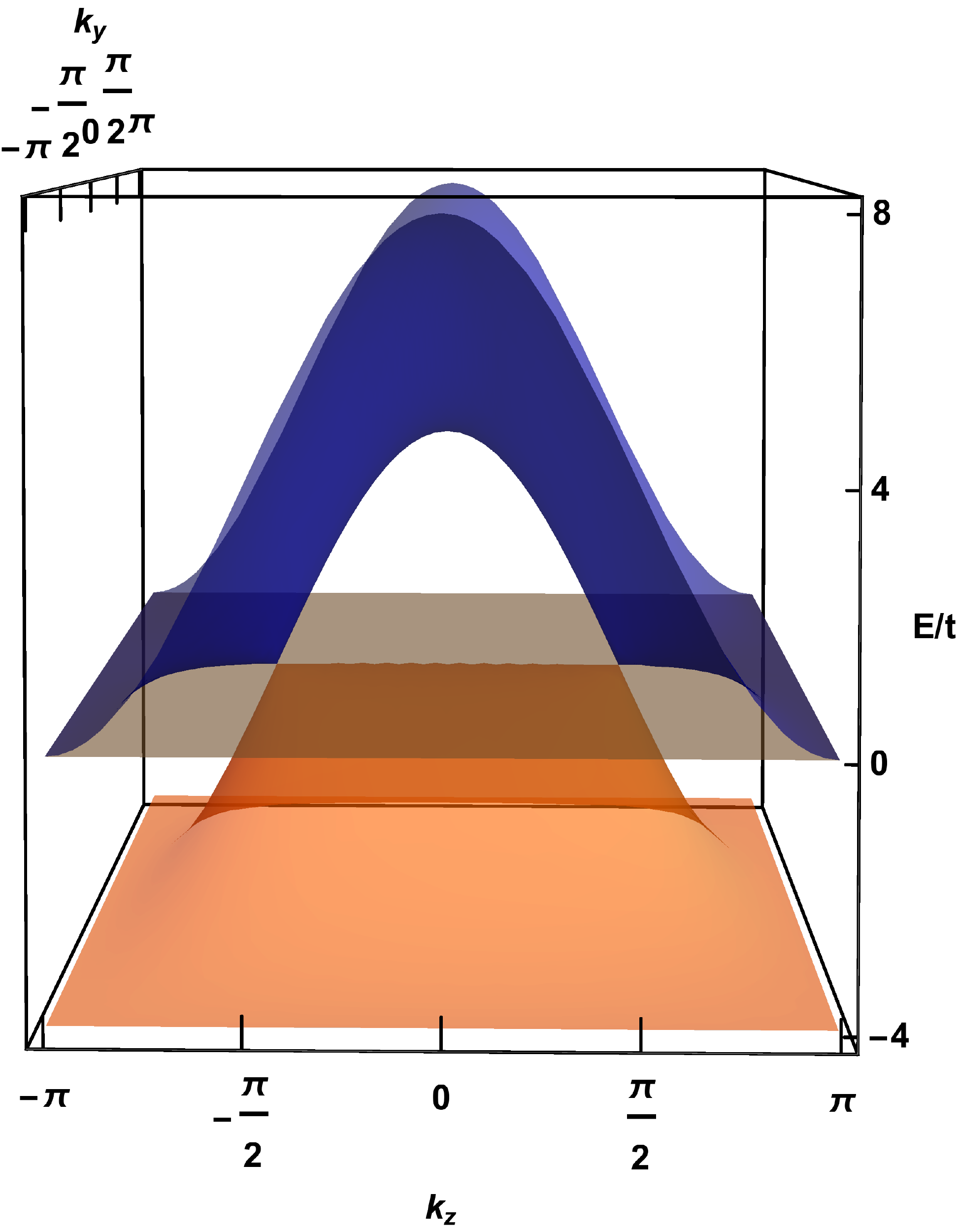}
\put(30,89){\begin{large} \bfseries $( \bf{c} )$ \end{large}}
\end{overpic}
\begin{overpic}[width=0.24\linewidth]{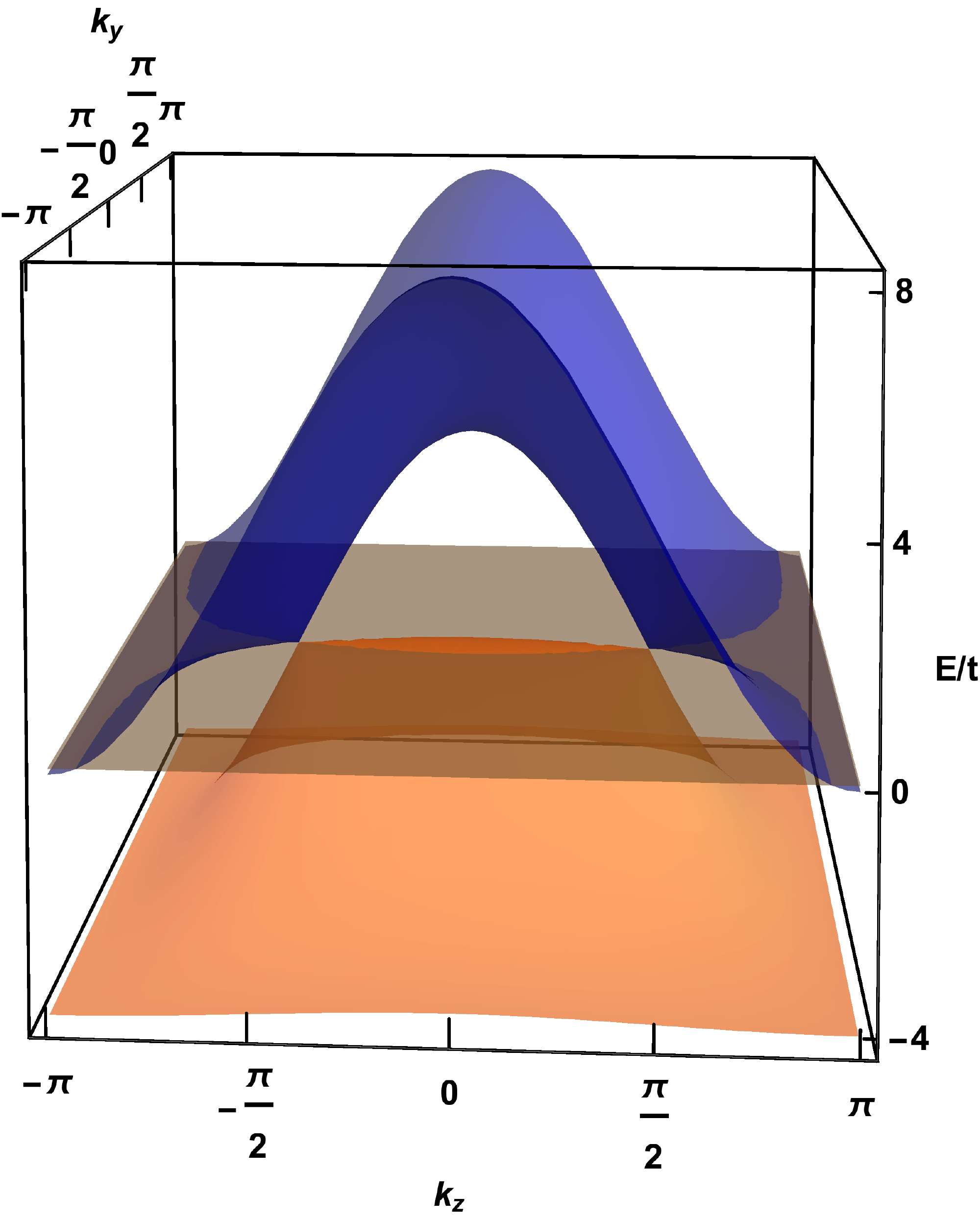}
\put(31,92){\begin{large} \bfseries $( \bf{d} )$ \end{large}}
\end{overpic}
		\includegraphics[width=0.24\linewidth]{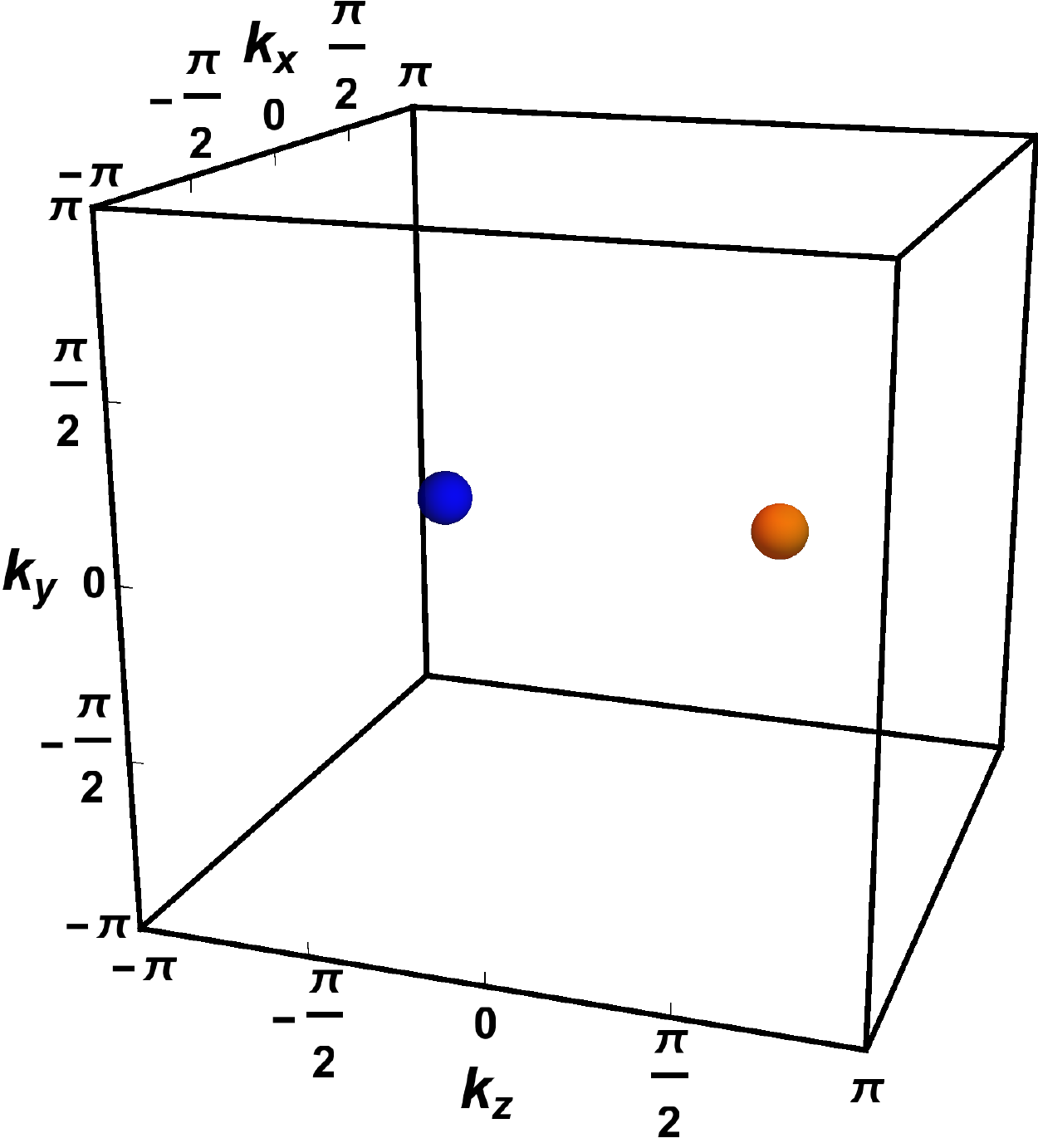}	%
		\includegraphics[width=0.24\linewidth]{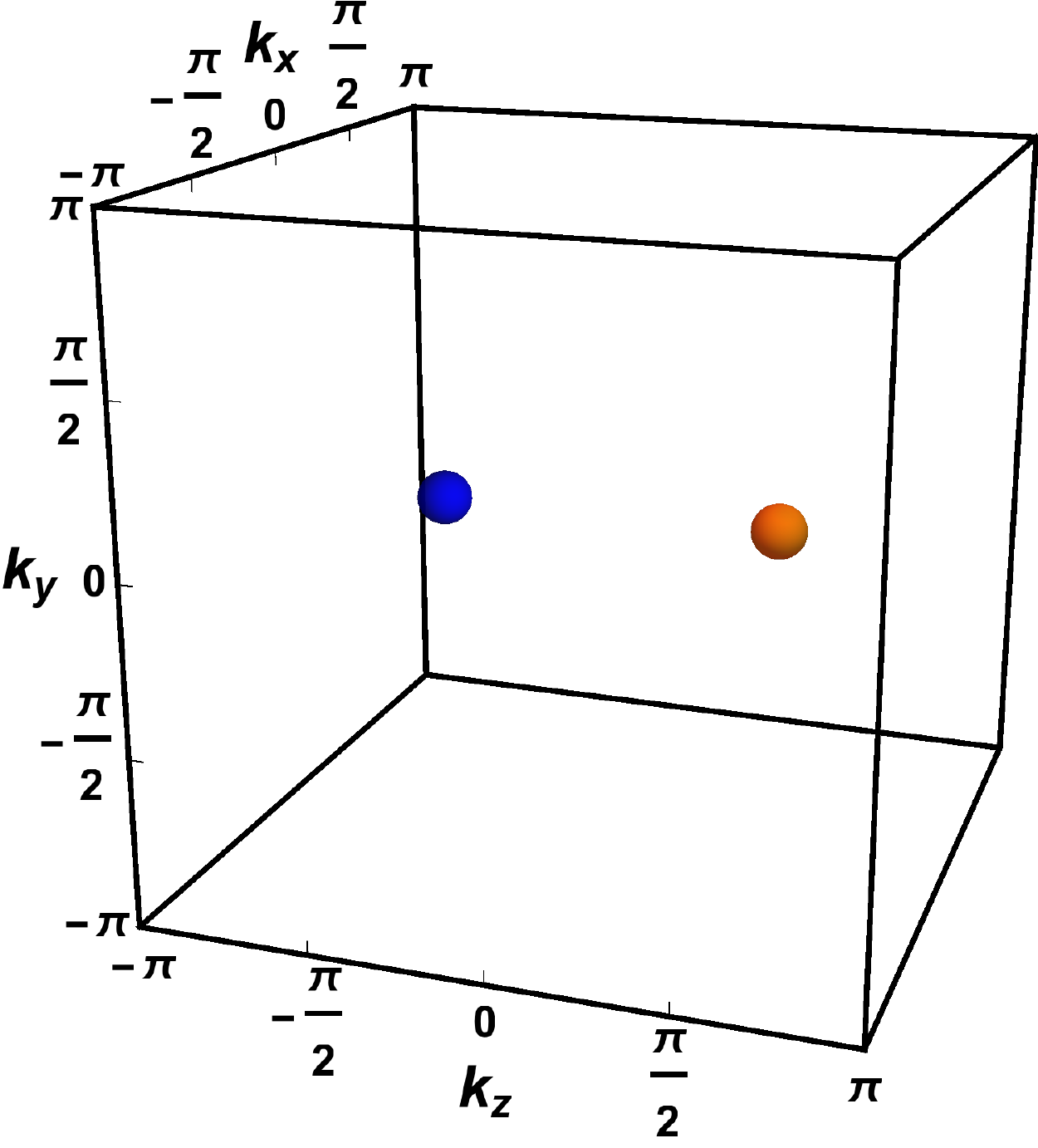}	
		\includegraphics[width=0.24\linewidth]{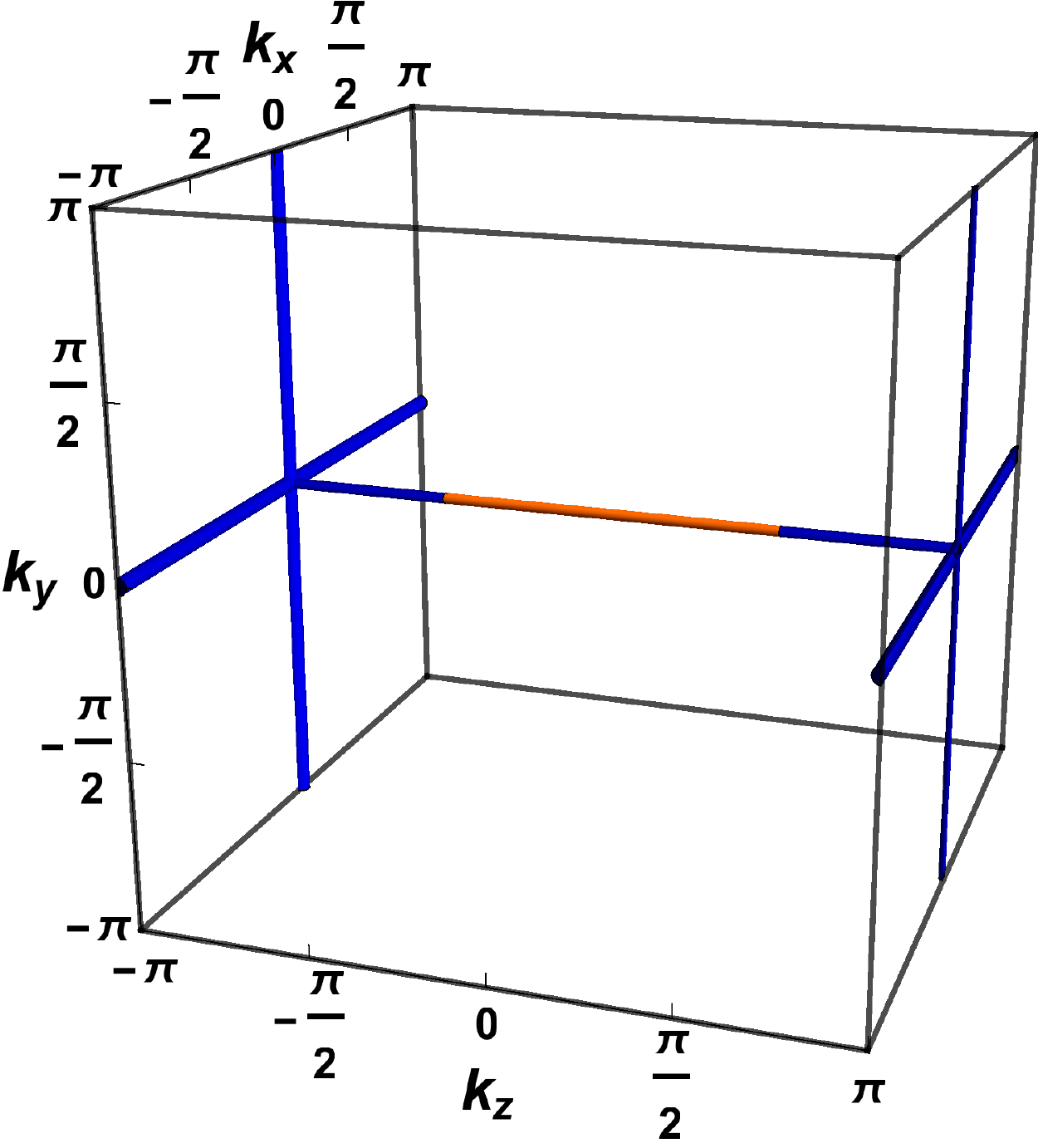}	
		\includegraphics[width=0.24\linewidth]{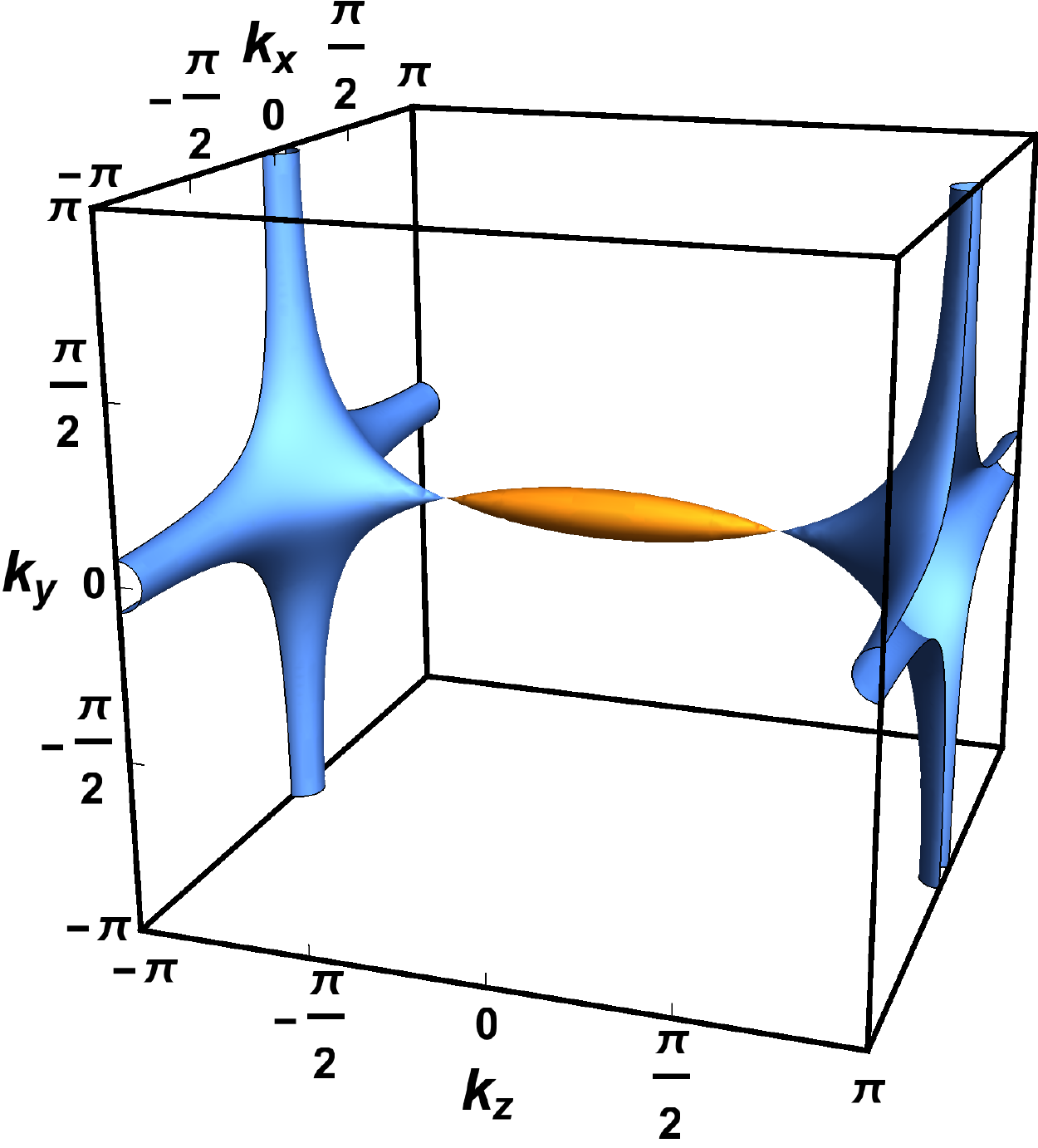}\\
		\includegraphics[width=0.99\linewidth]{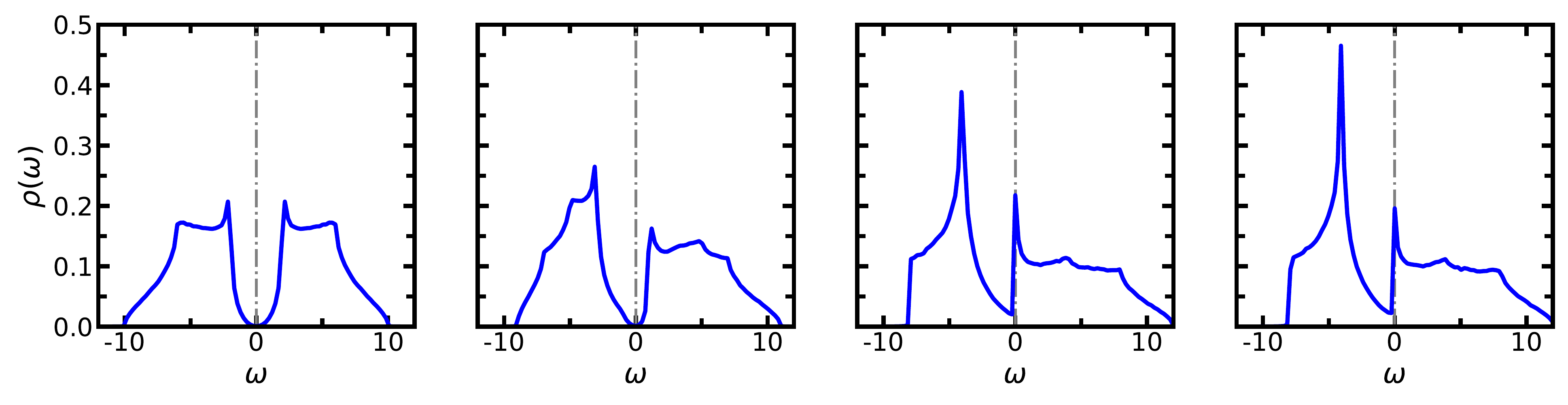}
	\vspace{-0.4cm}
	\caption{Figures labeled from column (a) to (d) are for tilting parameter $\gamma=0,1,2,2.1$,
 respectively. The upper row displays the energy spectrum using a cut geometry at $k_z=0$. The blue(orange) color denotes the conductance(valence) band, with the gray plane as the Fermi energy. The middle row shows the different topological configurations of Fermi surfaces, consistent with those in the first row. The type-I WSM at $\gamma=0$ and $\gamma =1$  
 (the left two: (a) and (b)) host point-like Fermi surfaces, the critical case at $\gamma=2$ (the middle (c)) exhibits as flat-line network Fermi surface and type-II WSM at $\gamma=2.1$ (on the right: (d)) has electron(blue) and hole(orange) Fermi pockets. The lower row displays the corresponding density of states from type-I to type-II WSM with the dashed gray line indicating the Fermi energy.   
	} \label{fig1}
\end{figure*}

\section{HAMILTONIAN MODEL}
\label{Model}

We start with a minimal two-band model with spin-orbit coupling on a cubic lattice considered in Ref. \cite{McCormick2017}, which hosts a pair of Weyl points with opposite chiralities in the Brillouin zone (BZ). The lattice constant is set as $a=1$. The bare Hamiltonian in momentum space is written as
\begin{equation}
\begin{split}
\mathcal{H}_{0}(\bm{k}) & = \gamma\left[\cos k_z-\cos k_{z0}\right] \sigma_{0}\\
&- \{ m\left[ 2-\cos k_x-\cos k_y \right]\\
&+2 t_z \left[\cos k_z-\cos k_{z0} \right] \}\sigma_{z}\\
&-2 t\sin k_x  \sigma_{x}-2 t\sin k_y  \sigma_{y}-\mu\sigma_0\, ,
\end{split}
\label{hamil}
\end{equation}
where $\sigma_{0}$ and $\sigma_{x,y,z}$ are the $2 \times 2$-identity matrix and Pauli matrices.  $\mu$ is the chemical potential of the system. The parameters $m$, $t_z$, and $t$ denote Zeeman term and SOC strengths. The $\gamma$-term has the effect of tilting the Weyl cones in the $k_z$-energy plane and is an even function of $k_z$. The constant parameter $k_{z0}$ determines the locations of the two Weyl nodes in 3D BZ. 

It is easy to verify that Hamiltonian in Eq. (\ref{hamil}) breaks time-reversal symmetry but keeps inversion symmetry and $\mathcal{C}_{4}$ rotation symmetry in the $x$-$y$ plane with respect to $z$ axis,
\begin{equation}
\begin{split}
\hat{\mathcal{P}}^{\dagger} \mathcal{H}_0(\bm{k}) \hat{\mathcal{P}} & =\mathcal{H}_0(\bm{-k}),\hat{\mathcal{P}}\leftrightarrow\sigma_{z}\\
\hat{\mathcal{T}}^{\dagger} \mathcal{H}_0(\bm{k}) \hat{\mathcal{T}} & \neq\mathcal{H}_0(\bm{-k}),\hat{\mathcal{T}}\leftrightarrow \mathcal{U} \mathcal{K}\\
\hat{\mathcal{R}}^{\dagger}(\theta)\mathcal{H}_0(\bm{k}) \hat{\mathcal{R}}(\theta) =& \mathcal{H}_0(\bm{k}), \hat{\mathcal{R}}(\theta) \leftrightarrow\hat{\mathcal{R}}_{k_z}(\theta)\hat{\mathcal{R}}_{\sigma_{z}}(\theta)
\end{split}
\label{sym}   
\end{equation}
with $\mathcal{U}$ a unitary matrix for a spin system and $\mathcal{K}$ a complex conjugate operator.  $\hat{\mathcal{R}}_{k_z}(\theta)=\begin{bmatrix} 
\cos(\theta) & \sin(\theta) \\
-\sin(\theta) & \cos(\theta) \\
\end{bmatrix}$ and $\hat{\mathcal{R}}_{\sigma_{3}}(\theta)=\begin{bmatrix} 
e^{i\frac{\theta}{2}} & 0\\
0 & e^{-i\frac{\theta}{2}}\\
\end{bmatrix}$ are  $2\times 2$ rotational matrices of the momentum basis of $\begin{bmatrix} 
k_x & k_y \end{bmatrix}^T$ and the spin basis of $\begin{bmatrix} 
\sigma_{x} & \sigma_{y} \end{bmatrix}^T$ respectively. Here the rotation angle is $\theta=\frac{n\pi}{2}$ with $n=1,2,3$.

The energy spectrum of the Hamiltonian above  is solved as
\begin{equation}
E_\pm (\bm{k})=T(\bm{k})\pm U(\bm{k})-\mu,
\end{equation}
with
\begin{equation}
\begin{split}
T(\bm{k})& =\gamma\left[\cos k_z-\cos k_{z0} \right], \\
U(\bm{k})& =\lbrace[ m(2-\cos k_x - \cos k_y )+t_z( \cos k_z-\cos k_{z0} )]^{2}\\
 & +4t^{2} [\sin^{2} k_x+\sin^{2} k_y ]\rbrace^{\frac{1}{2}} \\
\end{split}
\label{energy}   
\end{equation}
in which 
$T(\bm{k}) $ tilts the Weyl cones acting as the kinetic energy and $U(\bm{k})$ acts as the potential.

In the following, we focus on the half filling case with $\mu=0$ and the Hamiltonian model always possesses a pair of Weyl nodes at $\bm{k}=(0,0,\pm k_{z0})$ when solving $E_\pm (\bm{k})=0$. The parameters are chosen as $m=2t$ and $t_z=t$  with $t$ as energy unit and  $k_{z0}= \frac{\pi}{2}$ determine the position of Weyl nodes at $(0,0,\pm \frac{\pi}{2})$. The effective low-energy limit around two Weyl nodes reads as
\begin{equation}
\mathcal{H}_{0}(\bm{q}) = -\gamma \eta q_z \sigma_{0} +
2 t (\eta q_z \sigma_{z} -  q_x  \sigma_{x} - q_y  \sigma_{y})\, ,
\end{equation}
with $\eta=\sin (\pm k_{z0})$ that characterizes the opposite chirality($\pm 1$). According to the definitions of type-I and type-II WSM, if there exists a direction $\bm{e_k}$ in the Brillouin zone for $T(\bm{e_k})>U(\bm{e_k})$, one has a type-II WSM, and vice versa. For the Hamiltonian model in this paper, this direction is along the $z$-axis. When $\gamma <2 t_z$, the system belongs to type I (see Fig. \ref{fig1}(a) and (b)), and $\gamma >2 t_z$ turns into it a type-II (see Fig. \ref{fig1}(d)) with a critical point for the Lifshitz transition at $\gamma=2 t_z$ (see Fig. \ref{fig1}(c)).   

In order to study how electronic correlation effects make an impact on type-I and type-II WSM, we take an onsite Hubbard repulsion as the interaction term. In $\bm{k}$-space it is given by
\begin{equation}
\mathcal{H}_{int}=\frac{U}{N}\sum_{\bm{k},\bm{k}^{'},\bm{q}} c^{\dagger}_{\bm{k}\uparrow} c^{\dagger}_{\bm{k}^{'}\downarrow} c_{\bm{k}^{'}-\bm{q}\downarrow} c_{\bm{k}+\bm{q}\uparrow}.
\label{Hint}   
\end{equation}
This interaction term will be treated in the random-phase approximation to be discussed in the next section. 

\section{RPA METHOD WITH SPIN-ORBIT COUPLING}
\label{Method}
The Hamiltonian of WSM introduced above in Eq. (\ref{energy}) breaks SU(2) spin-rotational symmetry due to SOC. Accordingly, in the diagrammatic perturbative scheme that we do here, one has to take care of the spin indices. The generalized (spin-resolved) bare susceptibility in imaginary time-space is defined as
\begin{widetext}
\begin{equation}
\begin{split}
\chi^0_{\sigma_{1}\sigma_{2}\sigma_{3}\sigma_{4}}(\bm{q},\tau)&= \frac{1}{N}\sum_{\bm{k},\bm{k}^{'}}\left\langle T_{\tau}[c_{\bm{k}\sigma_{2}}^{\dagger}(\tau) c_{\bm{k}+\bm{q}\sigma_{3}}(\tau)c_{\bm{k}^{'}\sigma_{4}}^{\dagger}(0)c_{\bm{k}^{'}-\bm{q}\sigma_{1}}(0)]\right\rangle \\
&=-\frac{1}{N}\sum_{\bm{k}}G_{\sigma_1 \sigma_2}^{(0)}(\bm{k},-\tau) G_{\sigma_3 \sigma_4}^{(0)}(\bm{k}+\bm{q},\tau).
\end{split}
\end{equation}
\end{widetext}
with free but spin-dependent imaginary-time Green's function 
$G_{\sigma_1 \sigma_2}^{(0)}(\bm{k},\tau) $.
The transformation to Matsubara frequencies turns $\chi^0_{\sigma_{1}\sigma_{2}\sigma_{3}\sigma_{4}}(\bm{q},\tau)$
into $\chi^0_{\sigma_{1}\sigma_{2}\sigma_{3}\sigma_{4}}(\bm{q},i\nu_{n})$
\begin{equation}
\begin{split}
\chi^0_{\sigma_{1}\sigma_{2}\sigma_{3}\sigma_{4}}(\bm{q},i\nu_{n})&=- \frac{1}{N\beta}\sum_{\bm{k},m} 
G_{\sigma_1 \sigma_2}^{(0)}(\bm{k},i\omega_{m})\cdot \\&G_{\sigma_3 \sigma_4}^{(0)}(\bm{k}+\bm{q},i(\omega_{m}+\nu_{n})).
\end{split}
\end{equation}

For convenience, we perform a transformation of $\mathcal{H}_0(\bm{k})$ into its band basis with diagonal terms $\varepsilon_{i}(\bm{k})$ only. Then  $\chi^0_{\sigma_{1}\sigma_{2}\sigma_{3}\sigma_{4}}(\bm{q},i\omega_{n})$ in spin space can be written as
\begin{equation}
 \begin{split}
 \chi^0_{\sigma_{1}\sigma_{2}\sigma_{3}\sigma_{4}}(\bm{q},i\omega_{n})&=- \frac{1}{N}\sum_{\bm{k},i,j}
 \frac{n_{F}(\varepsilon_{i}(\bm{k}))-n_{F}(\varepsilon_{j}(\bm{k}+\bm{q}))}{\varepsilon_{i}(\bm{k})-\varepsilon_{j}(\bm{k}+\bm{q})+i\omega_{n}}\\
 &u_{\sigma_{1},i}(\bm{k})u_{\sigma_{2},i}^{*}(\bm{k}) u_{\sigma_{3},j}(\bm{k}+\bm{q})u_{\sigma_{4},j}^{*}(\bm{k}+\bm{q}),
 \end{split}
 \end{equation}
where $u_{\sigma, i}(\bm{k})$ and $\varepsilon_{i}(\bm{k})$ are the $\sigma$-component of the  $i^{th}$ eigenvector and the $i$-th eigenvalue of $\mathcal{H}_0(\bm{k})$. $n_F(\varepsilon_{i})=\frac{1}{e^{\beta\varepsilon_{i}}+1}$ denotes Fermi-Dirac distribution function.

Now we can write the bare susceptibility in a $4\times4$ full matrix with the notation $(\sigma_{2},\sigma_{3})$ as rows and $(\sigma_{4},\sigma_{1})$ as columns in which $(\sigma_{2},\sigma_{3})$ and $(\sigma_{4},\sigma_{1})$ are in the order of $(\begin{array}{cccc}
\uparrow\uparrow & \downarrow\uparrow & \uparrow\downarrow & \downarrow\downarrow\end{array})$
\begin{equation}
\hat{\chi}^{0} (\bm{q},i\omega)=\left[
\begin{array}{cccc}
 	\chi^0_{\uparrow\uparrow\uparrow\uparrow} & \chi^0_{\uparrow\uparrow\uparrow\downarrow} & \chi^0_{\downarrow\uparrow\uparrow\uparrow} & \chi^0_{\downarrow\uparrow\uparrow\downarrow}\\
 	\chi^0_{\uparrow\downarrow\uparrow\uparrow} & \chi^0_{\uparrow\downarrow\uparrow\downarrow} & \chi^0_{\downarrow\downarrow\uparrow\uparrow} & \chi^0_{\downarrow\downarrow\uparrow\downarrow}\\
 	\chi^0_{\uparrow\uparrow\downarrow\uparrow} & \chi^0_{\uparrow\uparrow\downarrow\downarrow} & \chi^0_{\downarrow\uparrow\downarrow\uparrow} & \chi^0_{\downarrow\uparrow\downarrow\downarrow}\\
 	\chi^0_{\uparrow\downarrow\downarrow\uparrow} & \chi^0_{\uparrow\downarrow\downarrow\downarrow} & \chi^0_{\downarrow\downarrow\downarrow\uparrow} & \chi^0_{\downarrow\downarrow\downarrow\downarrow}
 \end{array}
 \right].
 \label{chi0}
\end{equation}

\begin{figure}[ht]
 	\begin{center}
 	 \includegraphics[width=1.\linewidth]{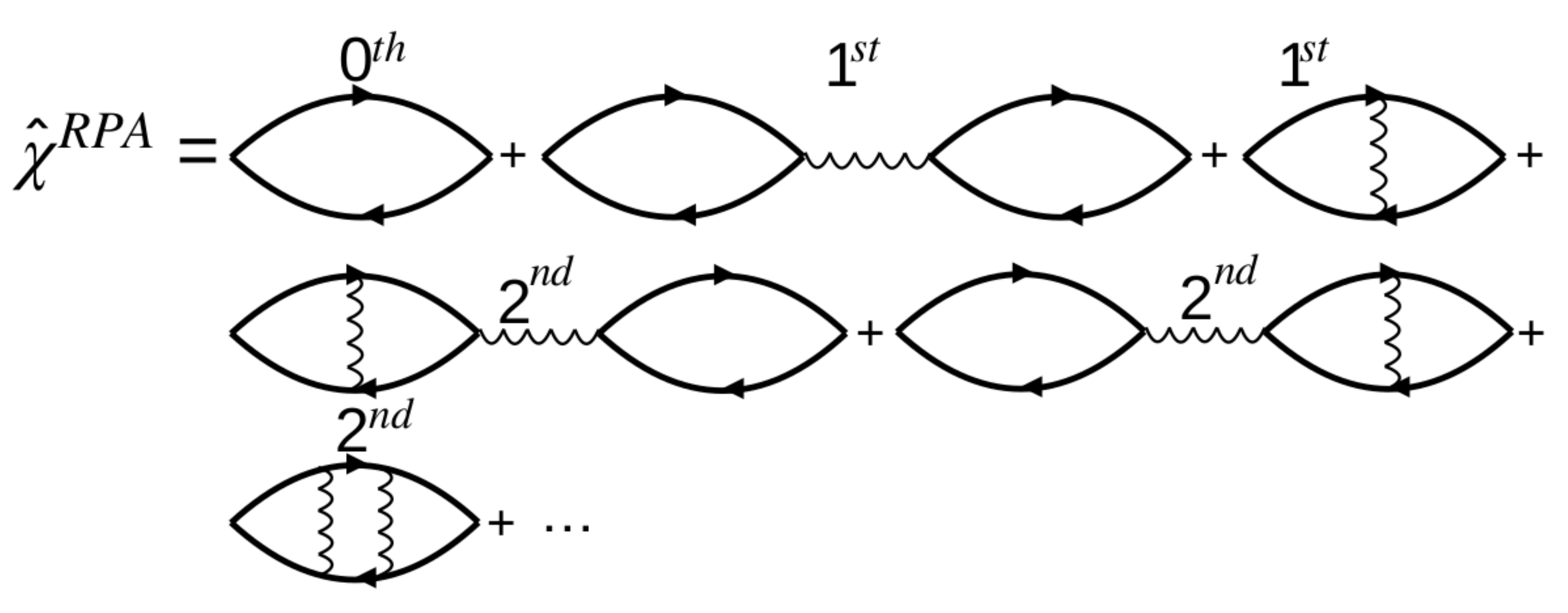}	
 	\end{center}
 	\vspace{-0.4cm}
 	\caption{The matrix-valued perturbation series $\hat{\chi}^{RPA}$ in Feynman diagram with all the possible mixtures of ladder and bubble diagram up to second order in $U$. Each solid line denotes a non-interacting Green's function $G_{\sigma_{i}\sigma_{j}}^{0}$ of a free particle. The wavy line represents the Hubbard onsite interaction matrix in Eq. (\ref{HubUMatrix}). It gives a nonzero contribution only when the spin index in conserved at both ends and opposite between the two ends of the wavy line.
 	} \label{fig2}
 \end{figure}
 
Thus the general standard expression of the susceptibility, which sums up all the infinite series of combinations of ladder and bubble Feynman diagrams, is expressed in a compact formula
\begin{equation}
\hat{\chi}^{RPA}(\bm{q},i\nu_{n})=\big[\mathds{1}+\hat{\chi}^{0}\cdot\hat{U}\big]^{-1}\hat{\chi}^{0},
\label{chiRPA}
\end{equation}
 where the matrix $\hat{U}$ is adopted as
\begin{equation} \label{HubUMatrix}
  \left[\begin{array}{cccc}
  0 & 0 & 0 & U\\
  0 & 0 & -U & 0\\
  0 & -U & 0 & 0\\
  U & 0 & 0 & 0
  \end{array}\right], 
\end{equation}
taking into account the spin structure of the onsite Hubbard interaction of Eq. (\ref{Hint}). 
\begin{figure*}[ht]
\begin{center}
 	 \includegraphics[width=1.0\linewidth]{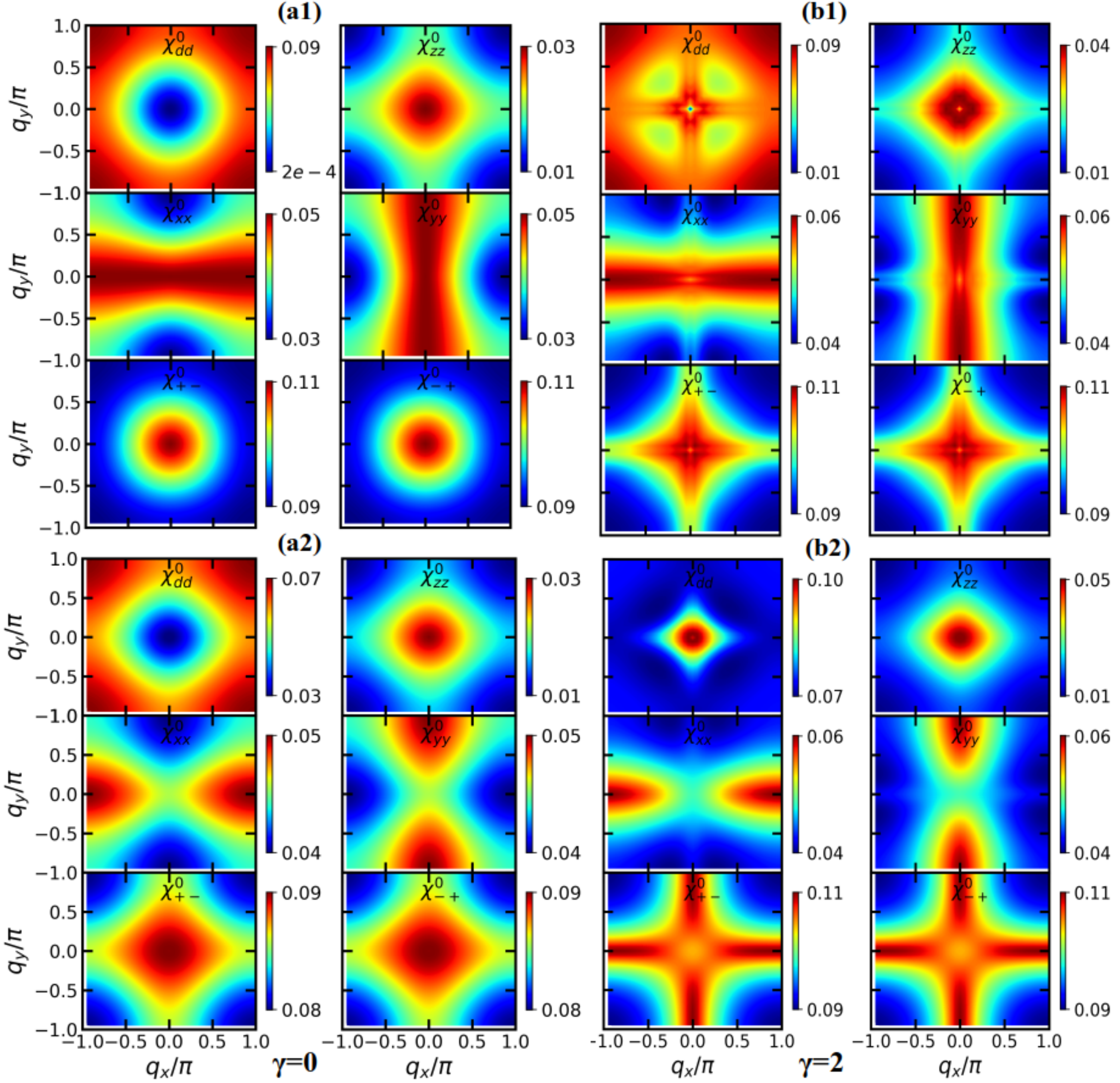}	
 	\end{center}
	\vspace{-0.4cm}
	\caption{Magnitudes of the real part of the bare susceptibilities as given in Eq. (\ref{chisus}) and as labeled in figures, displayed in a constant-$q_z$ plane. Left (a1/a2) panels: an ideal type-I WSM at $\gamma=0$. Right (b1/b2) panels: critical case of type-II WSM at $\gamma=2$. For (a1) and (b1) $q_z$ is fixed at $q_z=0$. (a2) and (b2) are for $q_z=\pi$.	
 } 
 \label{figchi0}
\end{figure*}

\begin{figure}[ht]
	\begin{center}
		\includegraphics[width=\linewidth]{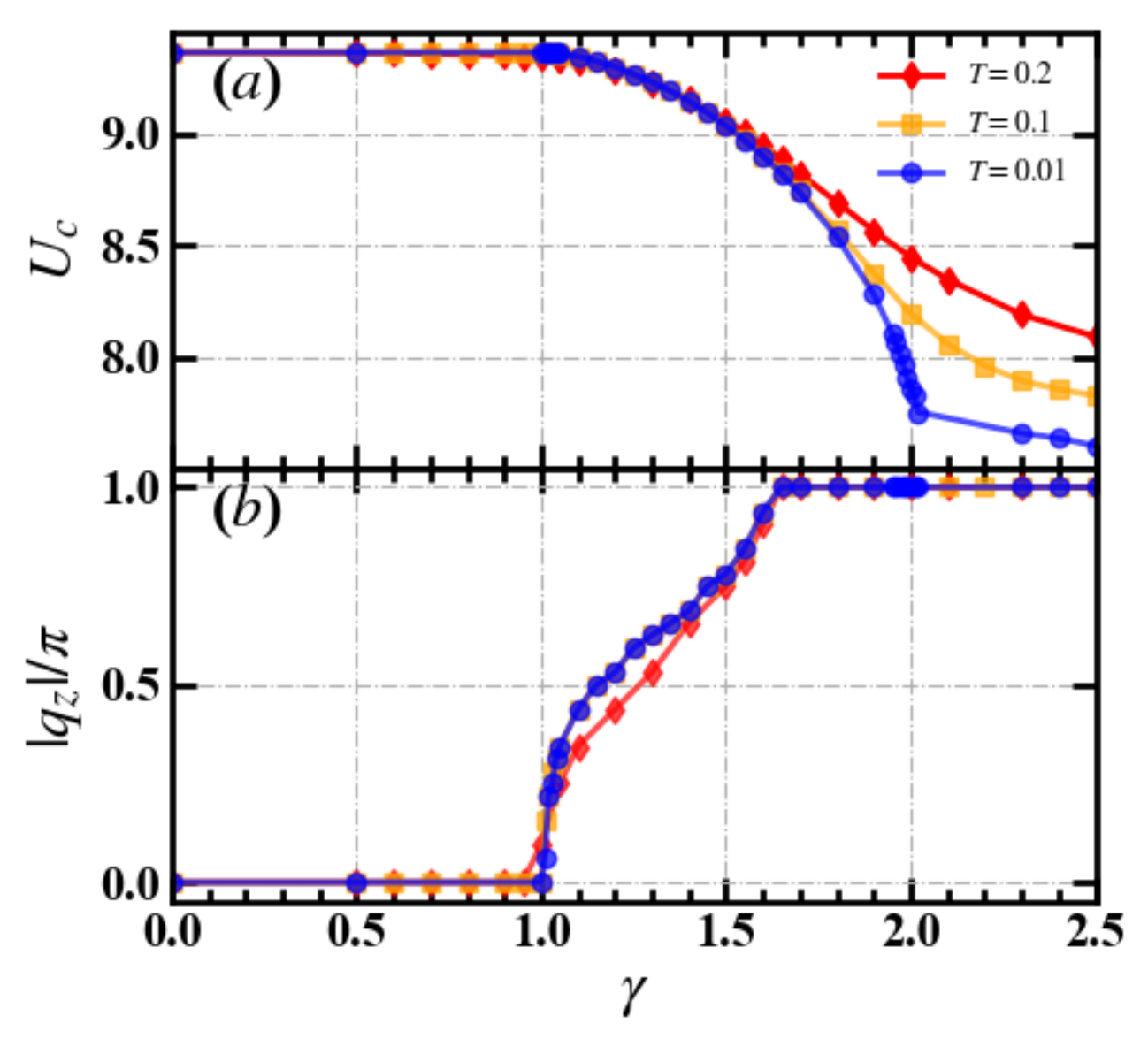}	
	\end{center}
    \vspace{-0.4cm}
	\caption{(a) Minimal critical values $U_c$ of the Hubbard interaction determined from Eq. (\ref{Uc}). (b) $q_z$-component of the  leading divergent transfer momentum $\bf{q}$  versus tilting parameter $\gamma$. Different colors from red to blue label different temperatures. 
	} \label{figUc}
\end{figure}

With the definitions of $\hat{\chi}^{0}$ and $\hat{\chi}^{RPA}$ above, the bare or RPA density-density correlation function ($\chi_{dd}^{0/RPA}$), longitudinal ($\chi_{zz}^{0/RPA}$) and transverse magnetic susceptibilities ($\chi_{+-(-+)}^{0/RPA}$ ) are given by
\begin{equation}
\begin{split}
\chi_{dd}^{0/RPA}(\bm{q},i\nu_{n})&=\chi_{\uparrow\uparrow\uparrow\uparrow}+\chi_{\uparrow\downarrow\downarrow\uparrow}+\chi_{\downarrow\uparrow\uparrow\downarrow}+\chi_{\downarrow\downarrow\downarrow\downarrow},\\
\chi_{zz}^{0/RPA}(\bm{q},i\nu_{n})&=\frac{1}{4}(\chi_{\uparrow\uparrow\uparrow\uparrow}-\chi_{\uparrow\downarrow\downarrow\uparrow}-\chi_{\downarrow\uparrow\uparrow\downarrow}+\chi_{\downarrow\downarrow\downarrow\downarrow}),\\
\chi_{+-(-+)}^{0/RPA}(\bm{q},i\nu_{n})&=\chi_{\uparrow\uparrow\downarrow\downarrow(\downarrow\downarrow\uparrow\uparrow)},
\end{split}
\label{chisus}
\end{equation}
where the superscript $^{0/RPA}$ has been omitted at the matrix components of $\hat{\chi}^{0}$ and $\hat{\chi}^{RPA}$, respectively, on the right hand sides. Besides, the planar spin susceptibilities $\chi_{xx(yy)}^{0/RPA}$ and $\chi_{+-(-+)}^{0/RPA}$ are related by
\begin{equation}
\begin{split}
\chi_{+-}^{0/RPA}(\bm{q},i\nu_{n})+\chi_{-+}^{0/RPA}(\bm{q},i\nu_{n})=\frac{1}{2}(\chi_{xx}^{0/RPA}+\chi_{yy}^{0/RPA}).
\end{split}
\end{equation}

\begin{figure}[ht]
	\begin{center}
		\includegraphics[width=0.75\linewidth]{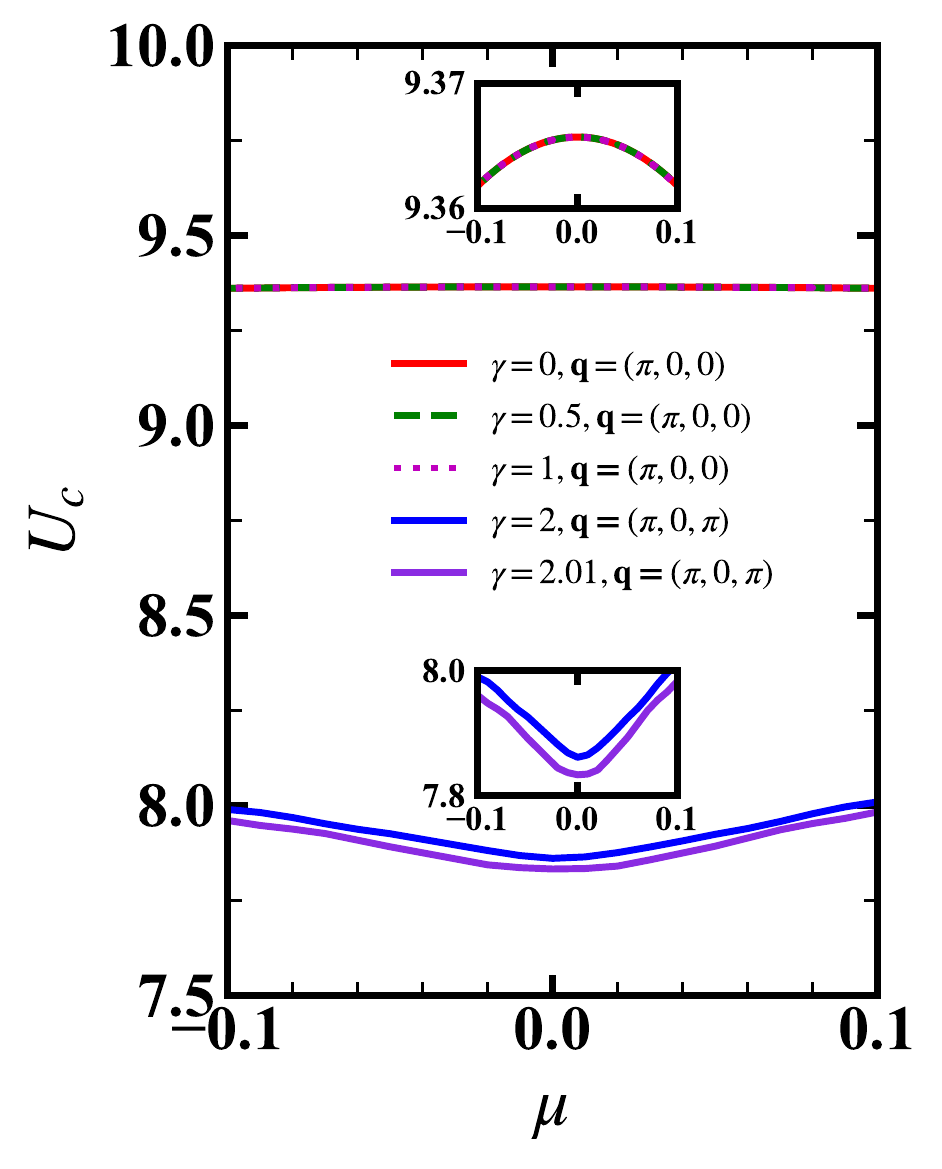}	
	\end{center}
	\vspace{-0.4cm}
	\caption{Critical $U_c$ of specific divergent magnetic orders depicted in Fig. \ref{figUc}(b) belonging to type-I and type-II WSM, as a function of a slightly doped chemical potential $\mu$ near the charge neutrality point. The upper(lower) inset displays the relative variation of $U_c$ at a finer scale.
	} \label{figchemicalUc}
\end{figure}

\section{NUMERICAL RESULTS FOR STATIC SUSCEPTIBILITIES}
\label{Result}

In this paper, we mainly study the static susceptibilities by concentrating on $\hat{\chi}^{RPA}(\bm{q},i\nu_{n}\rightarrow0)$.
By looking at the divergence of the matrix $\hat{\chi}^{RPA}$ described in Eq. (\ref{chiRPA}), two parts would possibly make contributions. One is from the divergence of the numerator of the bare susceptibilities $\hat{\chi}^{0}$ and the other is the occurrence of a zero determinant of the '\textit{denominator}' $\mathds{1}+\hat{\chi}^{0}\hat{U}$, where the onsite electronic interaction plays the role of control parameter. 

\subsection{Bare susceptibilities $\hat{\chi}^{0}$} \label{subsec:barechi}
We first check the real parts of the bare susceptibilities $\hat{\chi}^{0}$ defined in Eq. (\ref{chisus}). They display no signs of singularities for any momentum $\bf{q}$ in BZ. 
As two representative cases of those $\gamma=0$ (type-I) and $2$ (type-II) shown in Fig. \ref{figchi0}, as well as other parameter regime, the commonly response functions discussed above vary smoothly with comparable small fluctuations among each other. 
Note that due to the $\mathcal{C}_4$ rotational symmetry in $k_x$-$k_y$ plane of $\mathcal{H}_0$, the planar spin-spin susceptibilities still respect that symmetry, e.g., $\chi_{xx}^{0}(q_x,0,q_z)= \chi_{yy}^{0}(0,q_x,q_z)$. 
Compared the charge-density order $\chi^{0}_{dd}$ at $\bf{q}=0$ in Fig. \ref{figchi0}(a1) with that of $\bf{q}$ $=(0,0,\pi)$ in (a2) as well as Fig. \ref{figchi0}(b1) with (b2), the Weyl system is more in favor with an internode particle-hole excitation rather than an intranode one.
Especially, the charge-charge responses $\chi^{0}_{dd}$ of $\bf{q}=0$ are quite small quantities $\sim 10^{-4}$ for type-I WSM while enhance to the order $10^{-2}$ for type-II. 
This can be explained as for the type-I, the Fermi '$\textit{surface}$' are discrete Weyl nodes with a characteristic low-energy behavior of DOS $\varpropto\omega^{2}$ in the vicinity of a single Weyl node, vanishing as $\omega\rightarrow0$, as shown in Fig. \ref{fig1}(a)(b).
This leads to a small contribution of electron-hole pairs in the susceptibility.
For type-II at larger tilt $\gamma\ge 2$ there are peaks in the density of states near zero energy, as shown in Fig. \ref{fig1}(c)(d). Yet, they are one-sided, i.e., do not give rise to large particle-hole phase space at low energy when the peak is at the Fermi level.
Thus their electron-hole pair fluctuations improve but still remain at a finite value. 
From this point of view, the density responses at $\bf{q}=0$ towards electromagnetic fields could be used for reference in experiments to distinguish these two categories of WSMs. 

\subsection{Exotic spin-spin susceptibilities $\hat{\chi}^{RPA}$} 
\label{subsec:rpachi}

Next we turn to the RPA '\textit{denominator}' of Eq.  (\ref{chiRPA}) which contains the Hubbard interaction to search for potential symmetry broken phases. In generalization of the well-known Stoner argument, instabilities towards ordered phases occur for a divergence of the matrix inversion $[\mathds{1}+\hat{\chi}^{0}\cdot\hat{U}]^{-1}$. When we increase $U$ from zero, such a divergence occurs first where the most negative eigenvalue of matrix $\hat{\chi}^{0} \cdot \hat{U}$ becomes unity for a given wavevector $\bf q$. The $U$-value for this to happen defines a critical interaction strength 
\begin{equation}
U_c(\bf q)=-\frac{1}{\vert{\hat{\chi}^{0} (\bf q,0) \cdot\left[\begin{array}{cccc}
  0 & 0 & 0 & 1\\
  0 & 0 & -1 & 0\\
  0 & -1 & 0 & 0\\
  1 & 0 & 0 & 0
  \end{array}\right]}\vert_{\min}}.
\label{Uc}
\end{equation}
Below we will pick the minimal $U_c(\bf q)$ for all $\bf q$ as the relevant $U_c$ and additionally discuss at which wavevector this instability occurs.  
\begin{figure*}[ht]
\begin{overpic}[width=0.5\columnwidth]{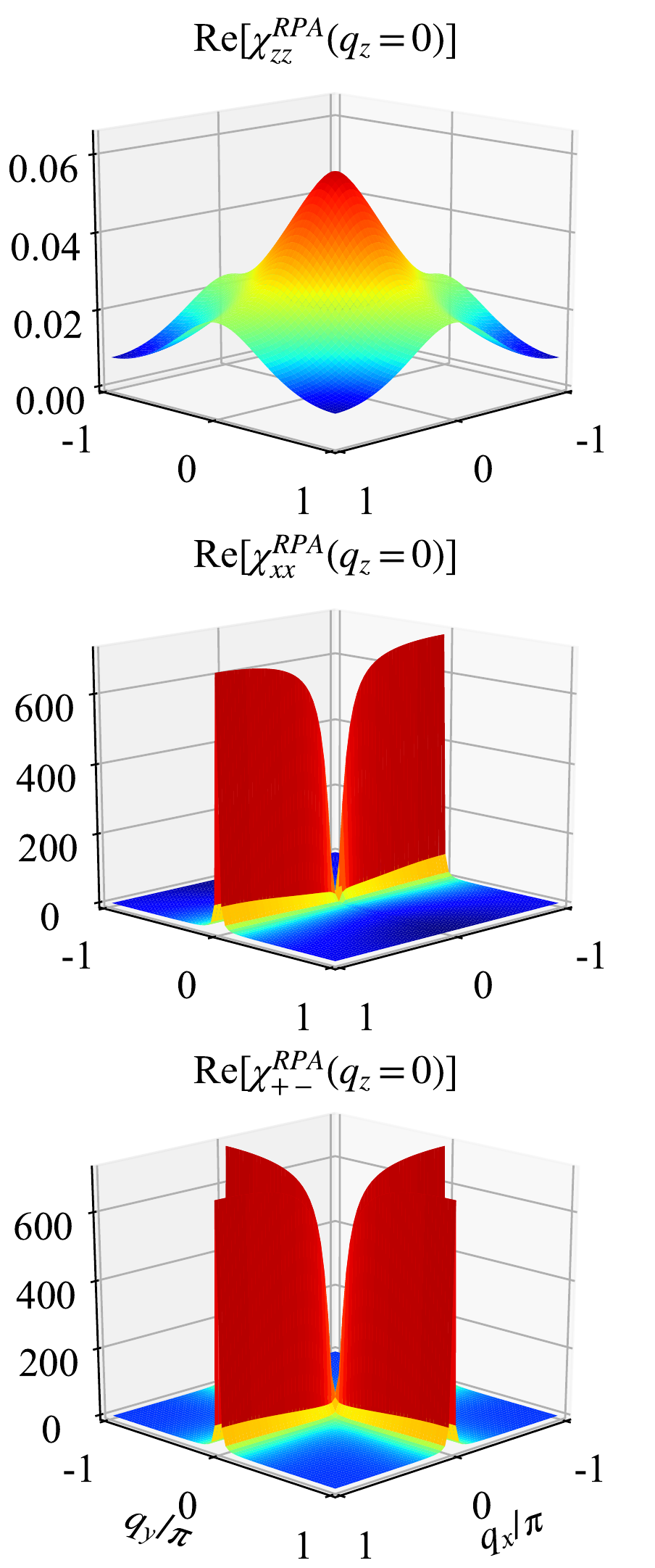}
\put(16,100){\begin{large} \bfseries $(\bf {a} )$ \end{large}}
\end{overpic}
\begin{overpic}[width=0.5\columnwidth]{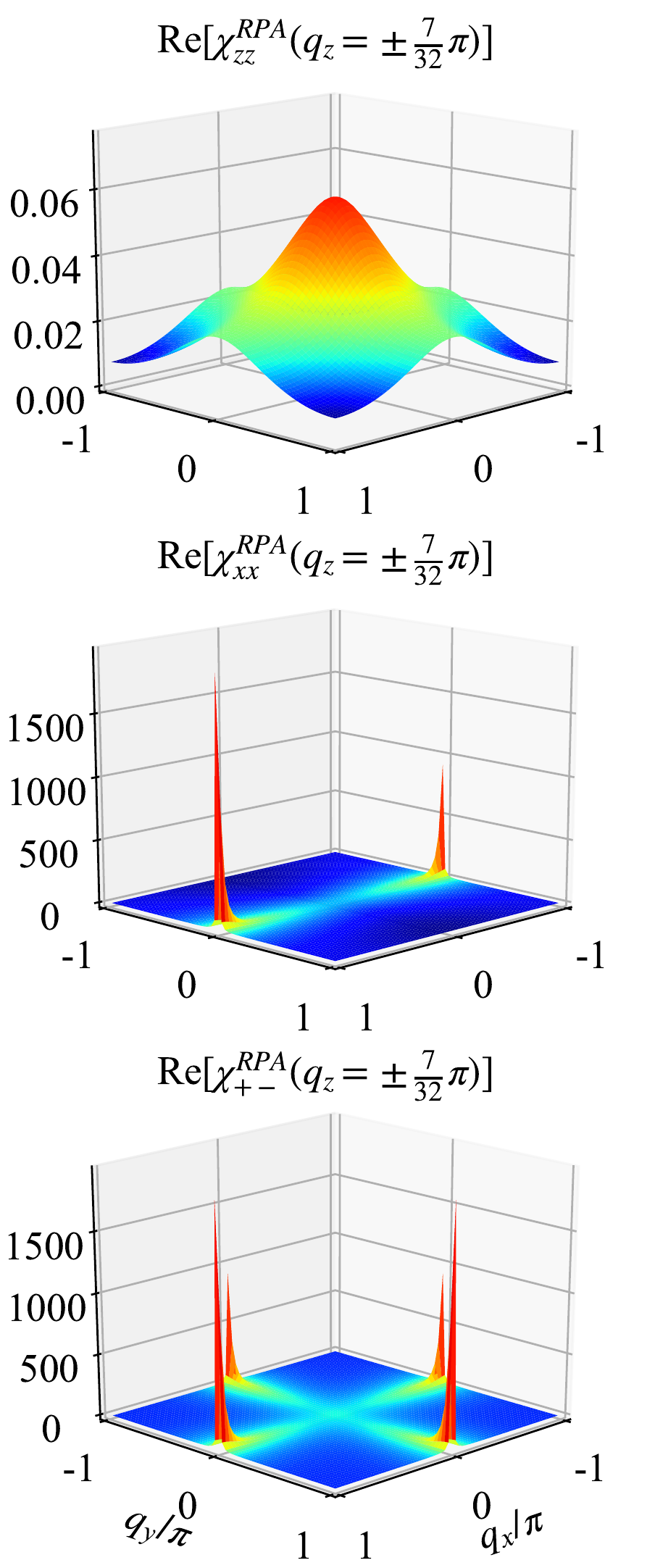}
\put(16,100){\begin{large} \bfseries $(\bf {b} )$\end{large}}
\end{overpic}
\begin{overpic}[width=0.5\columnwidth]{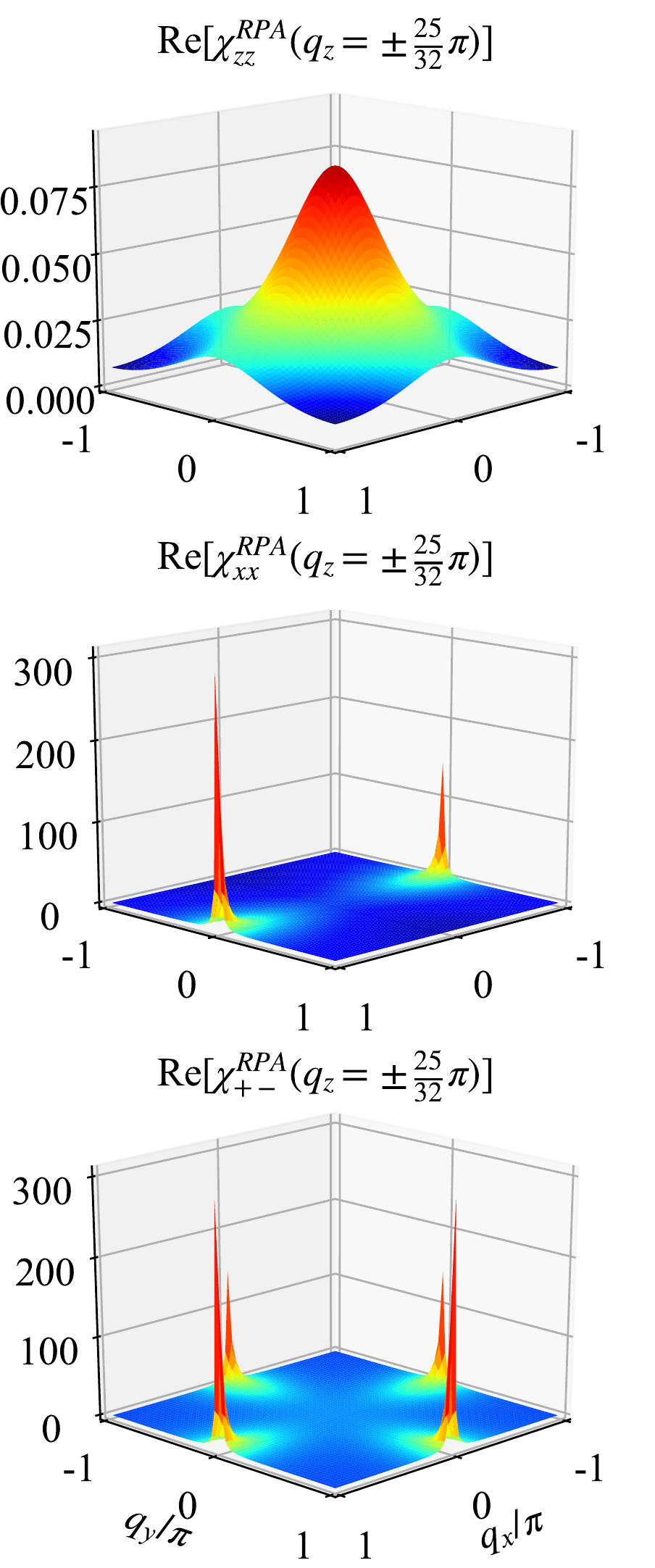}
\put(16,100){\begin{large} \bfseries $(\bf {c} )$ \end{large}}
\end{overpic}
\begin{overpic}[width=0.5\columnwidth]{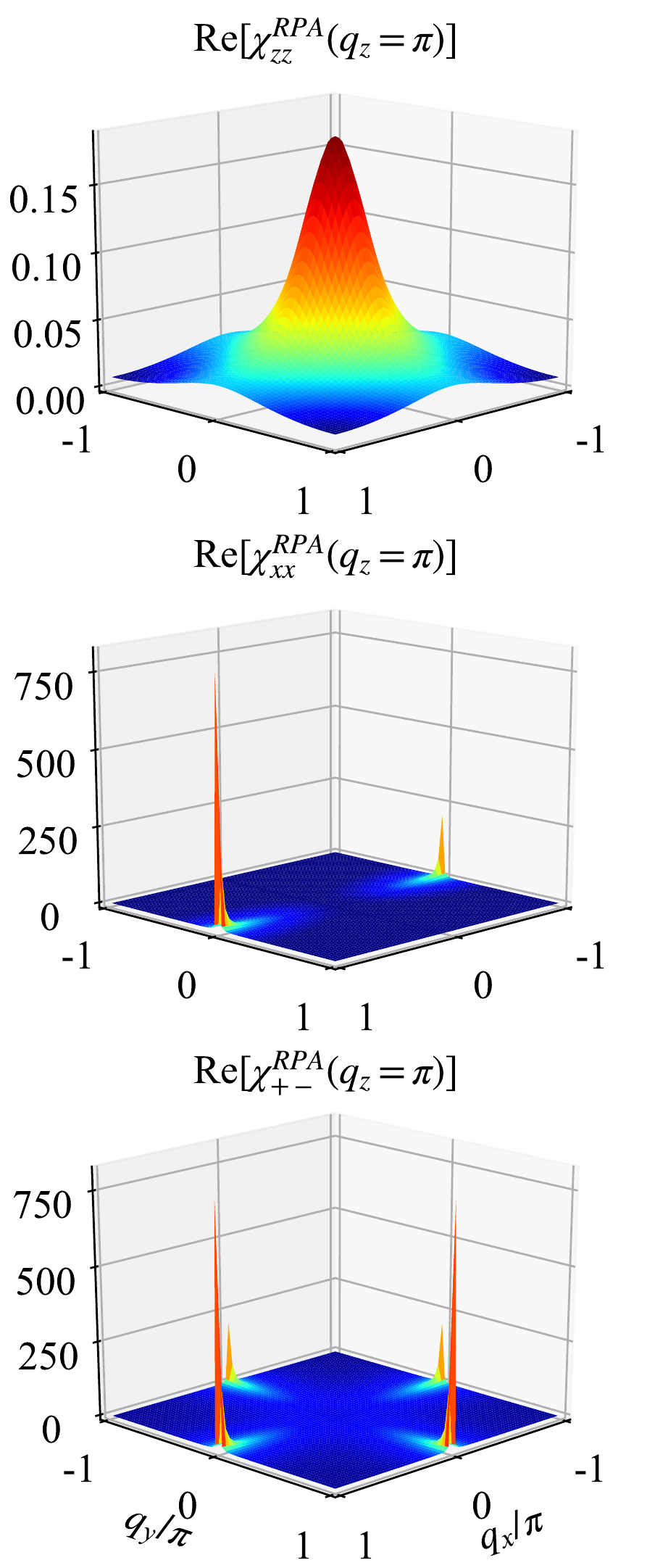}
\put(16,100){\begin{large} \bfseries $(\bf {d} )$ \end{large}}
\end{overpic}
	\vspace{-0.4cm}
	\caption{3D representations of the leading RPA spin density response in the planar momentum space $q_x$-$q_y$ for the Hubbard interaction $U$ approaching the corresponding critical $U_c$ in Fig. \ref{figUc}(a) from a smaller value. Panels (a) to (d) are for tilt parameter $\gamma=0,1.02,1.5,2$ with the $q_z$-components fixed at the locations of the peaks as displayed in Fig. \ref{figUc}(b).} 
 \label{figchiRPA}
\end{figure*}

In Fig. \ref{figUc} we show the numerical results for the $U_c$ values by searching through the whole BZ, obtained on momentum-$\bf{q}$ and -$\bf{k}$ meshes with an evenly distributive number $N=64^3$ in the 3D BZ. 
As displayed in Fig. \ref{figUc}(a), the critical $U_c \sim 0.5W_{\text{band}}$ decreases sharply towards the critical point $\gamma=2$, the transition point from type-I to type-II, which is consistent with the recent result obtained by DMFT method in an inversion symmetry broken model describing type-I and type-II WSM \cite{Kundu2021}.
The sharp feature at $\gamma =2$ at lower temperatures $T=0.01$ is increasingly smeared out with rising temperatures. Allowing for a slight electron(hole) doping near the nodal point Fermi energy, as shown in Fig. \ref{figchemicalUc}, the critical $U_c$ of the density order decreases compared to undoped case with $\delta^{2}\mu$ for type-I while increases linearly as $\delta\mu$ for type-II. 
The reason for a decrease for the type-I is that the doping near $\mu=0$ yields a finite density of single electron(hole) pocket at Fermi surface which in turn increases the particle-hole correlation slightly more than a pure nodal one. 
While for type-II, the missing of van Hove singularity at the Fermi surface with doping leads to a smaller density of electron-hole excitations and demands for a larger $U_c$. 
From this respect, the $U_c$ has an intimate relationship with the band structures and is sensitive to intrinsic changes of material.

Fig. \ref{figUc}(b) demonstrates that the component of the spin density order $\bf{q}$ along the direction of tilting direction($z$-axis) continuously increases from zero momentum condensation (ferromagnetic order) to large momentum-$\pi$ (antiferromagnetic order), in between undergoing incommensurate spin density orders, along with an increasing tilt term. 
The $q_z$ remains at zero with $\gamma$ in range of $\left[0,1\right]$ while  $U_c$ goes down slightly.
Within this interval for $\gamma <1$, they system remains in the same phase.
Beyond $\gamma=1$, $q_z$ starts to deviate from $0$, in which the magnetic order in $z$-direction forms spatial structure with a quasi-periodic modulation $\frac{2\pi}{q_z}$, and thus leads to an observable declination of the corresponding $U_c$. 
With increasing the tilt $\gamma$, ordering-$q_z$ for the instability reaches $\pi$ at $\gamma=1.65$ and after that remains stable.
The antiferromagnetic order of $q_z$ occurs at $\gamma=1.65$ before the of the original critical point $\gamma=2$ of the type-I to type-II transition.
Notice that our study does not imply a band structure renormalization by the interaction, which has been proved to shift the transition between type-I and type-II using Hartree-Fock method in Ref. \cite{Yixiang2017}.
Instead, our results show that features in the response functions do not necessarily coincide with qualitative changes of the band structure. 

 \begin{figure}[ht]
\begin{center}
	\includegraphics[width=\linewidth]{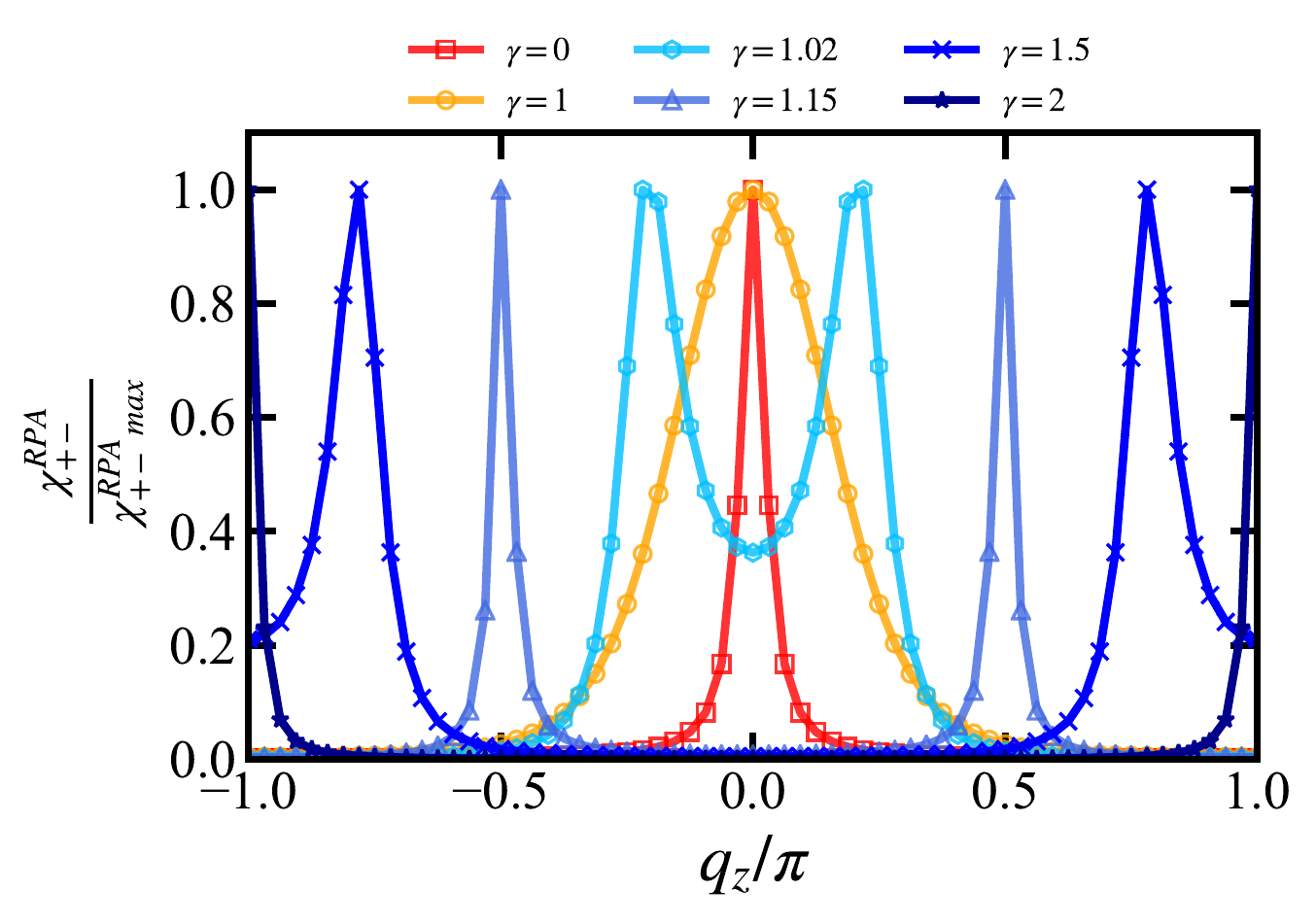}	
\end{center}
	\vspace{-0.4cm}
	\caption{The cuts of transverse spin susceptbility at $(0,\mathcal{\pi})$, normalized by the maximum value, versus $q_z$. Different markers are used for different parameters $\gamma$ of type-I and type-II WSMs.
	} \label{figsusqz}
\end{figure}

\begin{figure}[ht]      
\includegraphics[width=1.0\linewidth]{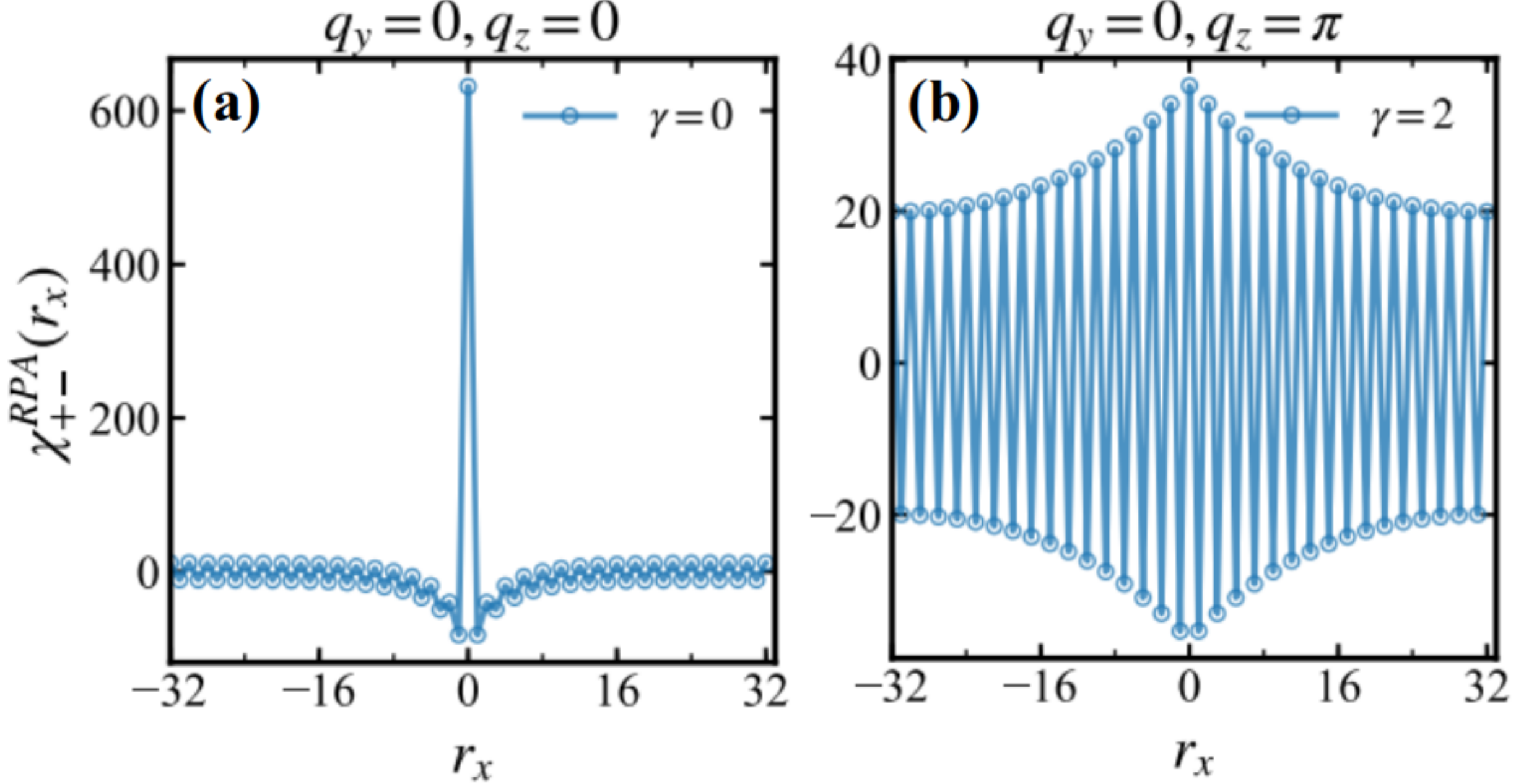}  
\begin{overpic}[width=0.3\linewidth]{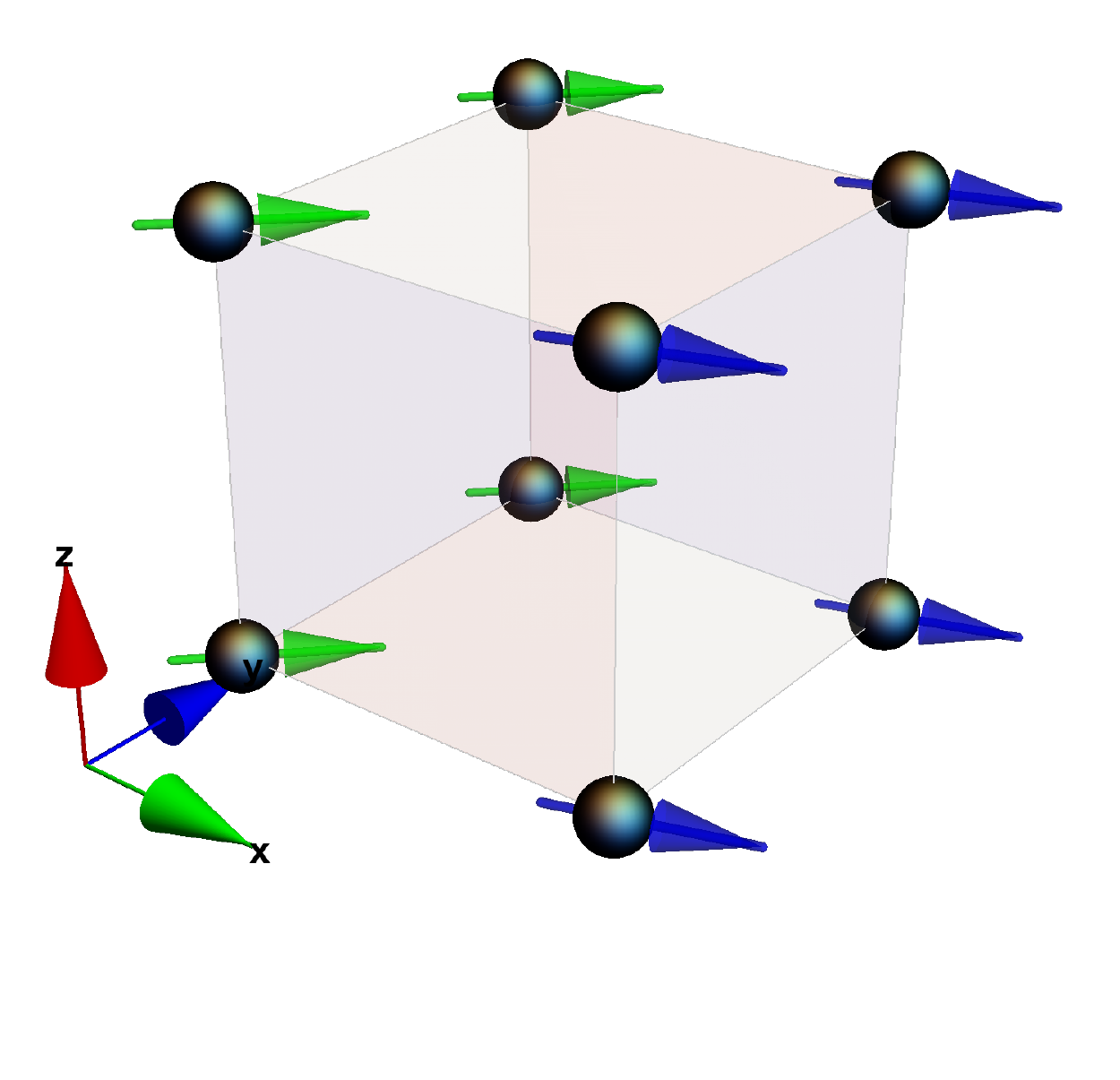}
\put(14,90){\begin{large}
\bfseries $(\bf {c})$\end{large} }
\end{overpic}
\begin{overpic}[width=0.32\linewidth]{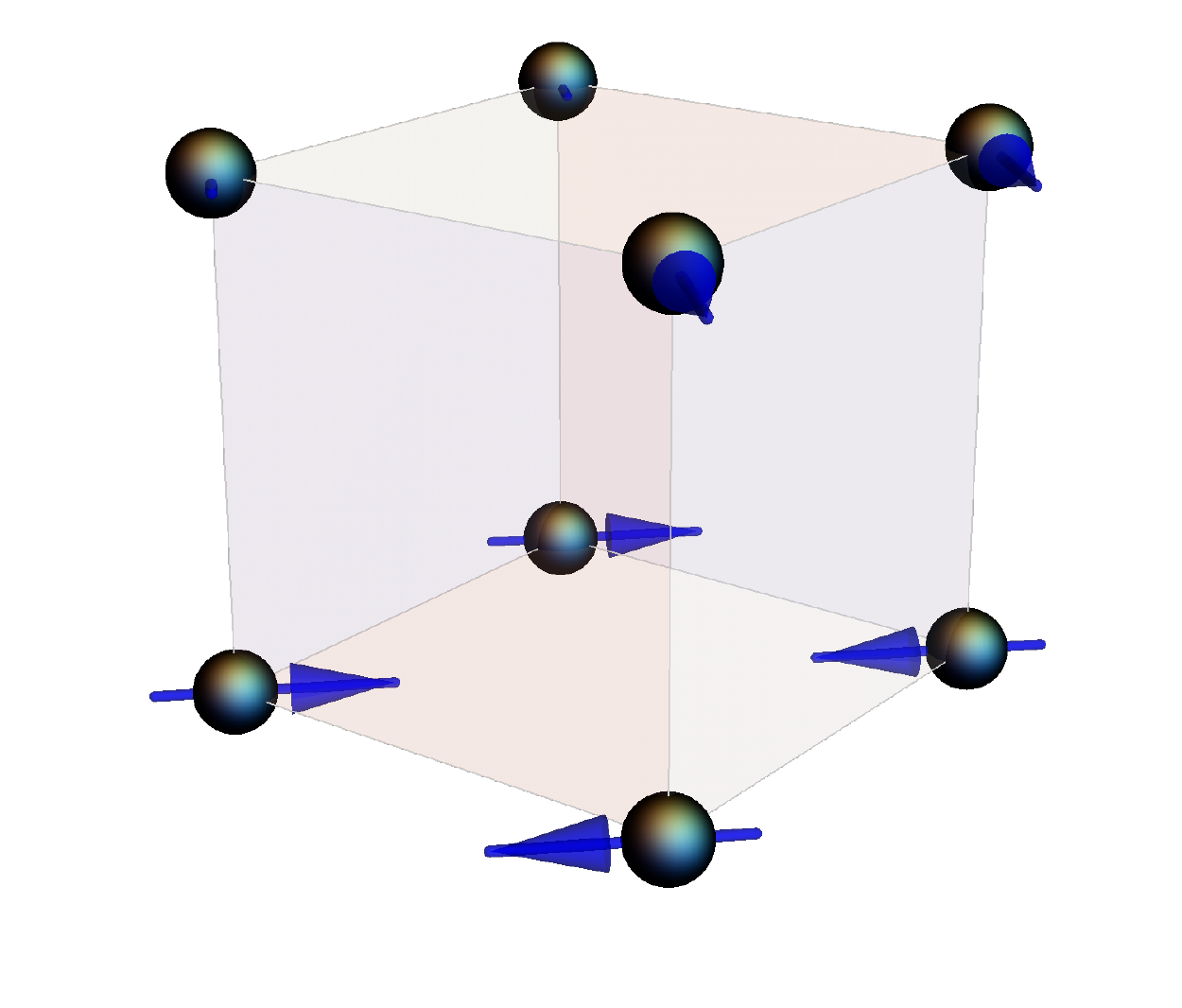}
\put(12,83){\begin{large} \bfseries $(\bf {d})$ \end{large}}
\end{overpic}
\begin{overpic}[width=0.32\linewidth]{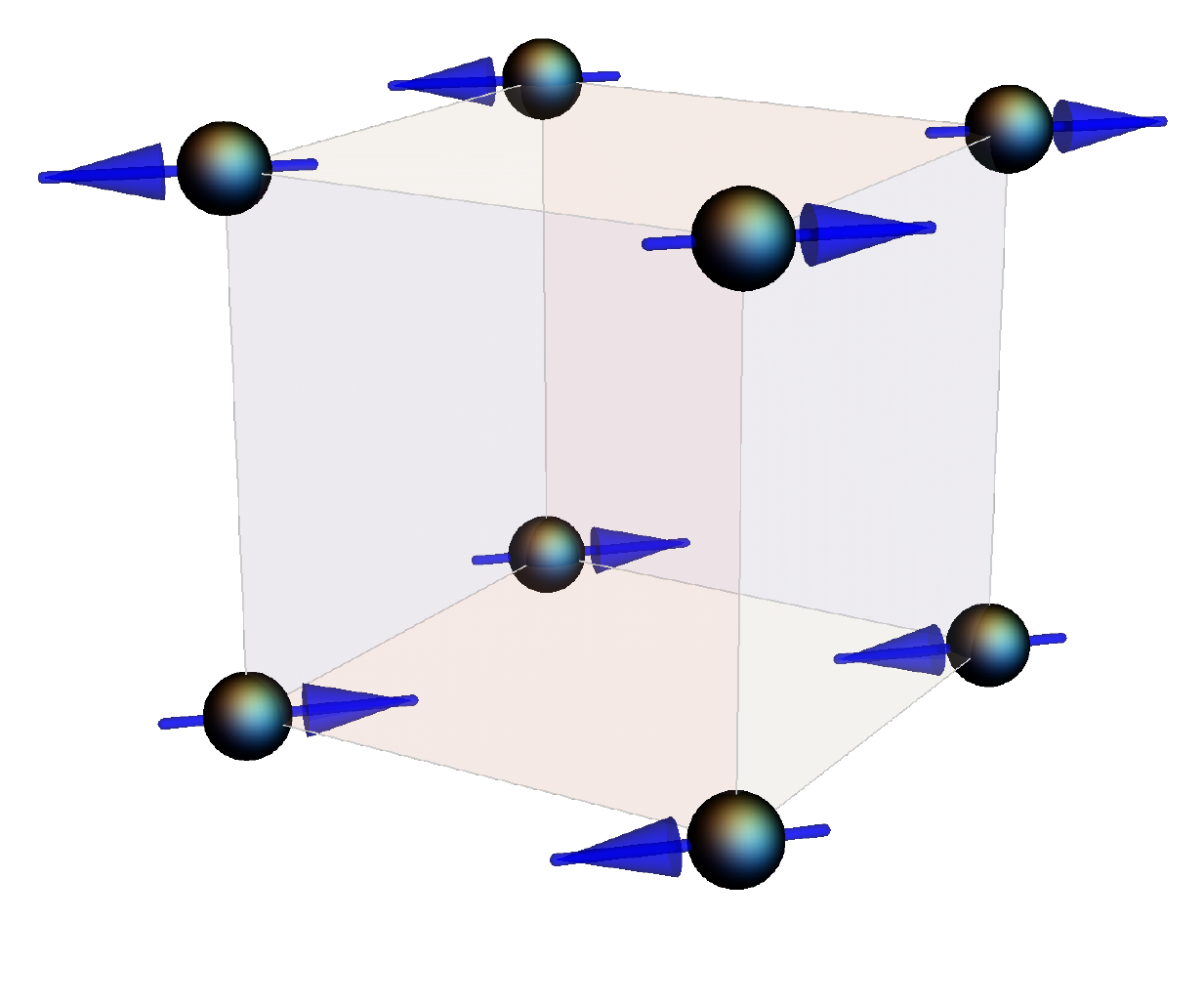}
\put(12,83){\begin{large} \bfseries $(\bf {e})$ \end{large}}
\end{overpic}
	\vspace{-0.4cm}
	\caption{ The upper row (a) and (b) are the Fourier transform of spin density order $\chi_{+-}^{RPA}(q_x)$ (shown in Fig. \ref{figchiRPA}(a) and (d)) to real space $\chi_{+-}^{RPA}(r_x)$ with a cut at $q_y=0$.  The lower panel (c)-(e) is an exhibition of emergent magnetism in 3D cubic lattice. Representative magnetic order $\bf{Q}_1$ (c): $(\{ q_x\neq0\},0,0)$ with ferromagnetism along $y$-axis(green arrows) independent of $x$-axis(blue arrows); (d): $(\pi,0,\frac{\pi}{4})$ with antiferromagnetism in $x$ direction and a spiral angle $\frac{\pi}{4}$ along $z$-axis; (e): $(\pi,0,\pi)$. Degenerate $\bf{Q}_2$ can be obtained by the $\mathcal{C}_{4z}$ symmetry correspondingly. Gray solid spheres denote the crystal lattices and colored arrows the net magnetism of a lattice.
	} \label{spin3D}
\end{figure}

Next, we visualize the most leading divergent channels in Fig. \ref{figchiRPA} in $q_x$-$q_y$ plane, in cuts at constant $q_z$s.
It is readily observed that under the short-ranged interaction the planar spin density susceptibility ($\chi^{RPA}_{+-}$) gets enhanced rather than the charge response ($\chi^{RPA}_{dd}$) and the longitudinal response ($\chi^{RPA}_{zz}$).
This indicates a strongly anisotropic response along the direction parallel and normal to the tilt orientation to external magnetic fields both in type-I and type-II WSMs. 
It is noted that the in-plane magnetic order still preserves the $\mathcal{C}_{4z}$ rotational crystal symmetry.
This leads to a double degeneracy of planar spin density wave $\chi^{RPA}_{+-(-+)}(q_x,0,q_z)=\chi^{RPA}_{+-(-+)}(0,q_x,q_z)$ when these susceptibilities diverge at $U_c$ with $\bf{q} \not= 0$. 

One interesting thing is now that for $\gamma=0$ in Fig. \ref{figchiRPA}(a), the susceptibility is close to its maximal value at $(\pi,0,0)$ or $(0,\pi,0)$ on a whole ridge for $(q_x,0,0)$ or $(0,q_y,0)$. In the extreme idealization, a susceptibility that diverges, e.g., at $(q_x,0,0)$ for all $q_x$ would feature layered ferromagnetic order in the $yz$-plane fully decoupled along the $x$-direction, as depicted in Fig. \ref{spin3D}(c). As there is some variation along these ridges, the decoupling of the layers is not perfect, instead the transformation of the ridge to real space shows a slight short-range anti-correlation. While we cannot firmly state that the ordered state will also be of the layer-liquid type, we definitely see a rather strong and exotic anisotropic spin correlation near the instability.   
A qualitatively similar picture applies to the rest of type-I cases $\gamma$ from 0 to 1 as well.
For a stronger tilt than $\gamma=1$, the planar spin response looses the ridges and  decays into a combination of two peaks at $(0, \pi, q_z)$ and $(\pi, 0, q_z)$, i.e., now with a modulation along the $z$-axis. 
The curves of the normalized transverse susceptibility in Fig. \ref{figsusqz} display how the spin density orders evolve along the tilt direction $z$ with increasing tilt term $\gamma$.
As $\gamma$ increases, it results in full range of magnetic textures including ferromagnetism, commensurate, incommensurate magnetism, and antiferromagnetism.
Especially when the tilt term exceeds $\gamma=1.65$,  it is antiferromagnetic or oppositely oriented in adjacent layers along the $z$ axis, as also shown in Fig. \ref{spin3D}(b) and (e). Here, all three space directions acquire long-range correlations.
Apart from these potential spontaneous symmetry breaking, the in-plane spin-density excitations are determined by the corresponding crystalline rotational symmetry of the model.

\section{SUMMARY}
\label{Summary}

We have studied a Hubbard model of type-I and type-II WSM ($\mathcal{T}$-reversal symmetry broken) governed by a tilting term. 
By the implementation of the RPA method for this non-SU(2)-invariant case, we explore the critical $U_c$ for magnetic or charge ordering at different tilt terms and temperatures, and the correspondent electromagnetic responses. 
By tuning the tilt term $\gamma$, temperature and doping chemical potential, $U_c$ decreases with more tilt and lower temperatures.
Upon doping away from half-filling $\mu=0$, type-I and type-II show different characteristics.
In fact, the $U_c$ reflect a dependence on the band structure near the Fermi surface. 
Regarding the main instabilities, spin density wave orders are favored over charge instabilities within the scope of RPA.
Besides, the Weyl systems behave anisotropically in spin and real space in the sense that they distinguish the direction of normal and parallel to the tilt $z$ axis in the reponse functions and instability tendencies.
The Hubbard interaction mainly drives planar spin density order with order parameter perpendicular to the $z$ direction  while longitudinal instabilities with spin order parameter along the tilt orientation are suppressed. 
Varying the tilt parameter $\gamma$, the moderately interacting type-I WSM first exhibits a 2D-layered magnetic configuration with a strongly anisotropic response in real space with almost decoupled ordering layers. Then, for larger $\gamma$ the model continues to develop into a 2D planar spin density wave instability, featuring in-plane combinations of $(\pi,0)$ and $(0,\pi)$ along with a translational-symmetry-broken order of a finite $q_z$ that modulates the 2D-layered sheet magnetism along the $z$-direction. 

As stated in the beginning, this RPA study may be viewed as a forerunner, e.g., to more sophisticated functional renormalization group studies of such systems. These will then also iron out on definite shortcoming of the present RPA study, the bias towards specific fluctuation channels and the lack of competition and interaction between different tendencies and fluctuations at different wavevectors. Nevertheless, renormalization group studies will become easier when based on the foundation of the present and comparable perturbative calculations that already lay out some main properties of the interaction physics. What also becomes clear from the RPA analysis is that this interaction physics is quite rich, with a sequence of potential orders and changes when the band structure is varied. A precise and controlled description of the interacting ground states may hence be a very good test-bed but also a challenge for many-body methods.

\begin{acknowledgments}
We thank Z.D. Yu, Y.C. He, J. Beyer, L. Klebl, Y.C. Liu and J. Ehrlich for discussions. The German Science Foundation (DFG) is acknowledged for support through RTG 1995, and RWTH-HPC for granting computing time.
\end{acknowledgments}

\bibliography{rpa.bib}
%------------------------------------------------
\end{document}